\documentclass[preprint,eqsecnum,floats,aps]{revtex4}
\usepackage{graphicx}
\usepackage{bm}
\usepackage{epsfig}
\usepackage{mathrsfs}
\begin{document}
\newcommand{\Mass}{\mathrm{M}}
\newcommand{\mass}{\mathrm{m}}
\def\Ss{\mathscr{S}}
\def\Ys{\mathscr{Y}}
\def\Os{\mathscr{O}}
\def\threej#1#2#3#4#5#6{\left(\negthinspace\begin{array}{ccc}
#1&#2&#3\\#4&#5&#6\end{array}\right)}
\def\sixj#1#2#3#4#5#6{\left\{\negthinspace\begin{array}{ccc}
#1&#2&#3\\#4&#5&#6\end{array}\right\}}
\def\up{u_{{\rm p}}}
\def\vp{v_{{\rm p}}}
\def\un{u_{{\rm n}}}
\def\vn{v_{{\rm n}}}
\def\upp{u_{p'}}
\def\vpp{v_{p'}}
\def\unp{u_{n'}}
\def\vnp{v_{n'}}
\def\vk{v_{{\rm k}}}
\def\uk{u_{{\rm k}}}
\def\ga{\overline{g}_{\mbox{\tiny A}}}
\def\gv{\overline{g}_{\mbox{\tiny V}}}
\def\gp{\overline{g}_{\mbox{\tiny P}}}
\def\gpa{\overline{g}_{\mbox{\tiny P1}}}
\def\gpb{\overline{g}_{\mbox{\tiny P2}}}
\def\gw{\overline{g}_{\mbox{\tiny W}}}
\def\gm{\overline{g}_{\mbox{\tiny M}}}
\def\gA{g_{\mbox{\tiny A}}}
\def\rA{r_{\mbox{\tiny A}}}
\def\gAeff{g_{\mbox{\tiny A,eff}}}
\def\gV{g_{\mbox{\tiny V}}}
\def\gM{g_{\mbox{\tiny M}}}
\def\gS{g_{\mbox{\tiny S}}}
\def\gT{g_{\mbox{\tiny T}}}
\def\gP{g_{\mbox{\tiny P}}}
\def\gPeff{g_{\mbox{\tiny P,eff}}}
\def\GA{G_{\mbox{\tiny A}}}
\def\GV{G_{\mbox{\tiny V}}}
\def\GF{G_{\mbox{\tiny F}}}
\def\lsim{\:\raisebox{-0.5ex}{$\stackrel{\textstyle<}{\sim}$}\:}
\def\gsim{\:\raisebox{-0.5ex}{$\stackrel{\textstyle>}{\sim}$}\:}
\def\be{\begin{equation}}
\def\ee{\end{equation}}
\def\br{\begin{eqnarray}}
\def\er{\end{eqnarray}}
\def\brn{\begin{eqnarray*}}
\def\ern{\end{eqnarray*}}
\def\E {{{\cal E}}}
\def\T {{{\cal T}}}
\def\ie{{\em i.e. }}
\def\nn{\nonumber }
\def\rf#1{{(\ref{#1})}}
\def\sss{\scriptscriptstyle}
\def\ss{\scriptstyle}
\def\eb {{\bf e}}
\def\kb {{\bf k}}
\def\lb {{\bf l}}
\def\Ib {{\bf I}}
\def\Ob{ {\bf O}}
\def\pb {{\bf p}}
\def\Pb {{\bf P}}
\def\qb {{\bf q}}
\def\qbh {\hat{\bf q}}
\def\rb {{\bf r}}
\def\vb {{\bf v}}
\def\xb {{\bf x}}
\def\mbn{\mbox{\boldmath$\nabla$}}
\def\sq{\sqrt{2}}
\def\a {{\alpha}}
\def\b {{\beta}}
\def\d{\dagger}
\def\e{\epsilon}
\def\g {{\gamma}}
\def\s {{\sigma}}
\def\k {{\kappa}}
\def\l {{\lambda}}
\def\t {{\tau}}
\def\go{\rightarrow}
\def\mbs{\mbox{\boldmath$\sigma$}}
\def\Jb{ {\bf J}}
\def\x{\times}
\def\ket#1{|#1 \rangle}
\def\bra#1{\langle #1|}
\def\Ket#1{||#1 \rangle}
\def\Bra#1{\langle #1||}
\def\ov#1#2{\langle #1 | #2  \rangle }
\def\M {{{\cal M}}}
\def\O {{{\cal O}}}
\def\rh {\hat{r}}
\def\kh {\hat{k}}
\def\zh {\hat{z}}
\def\threej#1#2#3#4#5#6{\left(\negthinspace\begin{array}{ccc}
#1&#2&#3\\#4&#5&#6\end{array}\right)}
\def\del {\delta}
\def\lra{\leftrightarrow}
\def\etal{{\it et al.}}
\def\kr {{\bf k}\cdot{\bf r}}
\def\sqi{\frac{1}{\sqrt{2}}}
\def\etc{ {\it etc}}
\def\w {{\omega}}
\def\bit{\begin{itemize}}
\def\eit{\end{itemize}}
\def\L {{{\cal L}}}
\title{Neutrino-Nucleus Reactions and Muon Capture in $^{12}C$ }
\author{ F. Krmpoti\'c$^{1,2,3}$,   A. Samana$^{1}$ and  A. Mariano$^{2,4}$
 }
\affiliation{$^1$Instituto de F\'{\i}sica, Universidade de S\~ao Paulo,
05315-970 S\~ao Paulo-SP, Brazil}
\affiliation{$^2$Instituto de F\'{\i}sica La Plata,
CONICET, 1900 La Plata, Argentina}
\affiliation{$^3$Facultad de Ciencias Astron\'omicas y Geof\'{\i}sicas, 
Universidad Nacional de La Plata, 1900 La Plata, Argentina,}
\affiliation{$^4$Departamento de F\'{\i}sica, Universidad Nacional de
La Plata,\\C. C. 67, 1900 La Plata, Argentina}

\begin{abstract}

The neutrino-nucleus cross section and the muon capture rate
are discussed within a simple formalism which  facilitates
the nuclear structure calculations.
The corresponding formulae only depend on four types of nuclear 
matrix elements, which are currently used in the nuclear beta decay.
We have also considered
the non-locality effects arising from the  velocity-dependent terms in
the hadronic current.
We show that for both observables in $^{12}C$  the higher order
relativistic corrections are of the order of $\sim 5\%$ only, and
therefore do not play a significant role.
As nuclear model framework we use the projected QRPA (PQRPA) and show
that the number projection plays a crucial role in removing the degeneracy
between the proton-neutron two quasiparticle states  at the level of the mean
field.
Comparison is done with both the experimental data and 
the previous shell model calculations. Possible consequences of the
present study on the determination of the $\nu_\mu\go\nu_e$
neutrino oscillation probability are briefly addressed.
\vspace{2.6cm}

{\it PACS number:} 23.40.-s,  23.40.Bw,  23.40.Hc,  25.30.Pt, 21.60.Jz

\vspace{.5cm}

{\it Keywords:} neutrino-nucleus cross section, muon capture, projected
QRPA
\end{abstract}

\maketitle

\newpage

\section{Introduction}
\label{sec1}
The  semileptonic weak interactions with nuclei include a rich
variety of processes, such as the neutrino (antineutrino)
scattering, charged lepton capture, $e^{\pm}$ decays, etc, and we
have at our disposal the results of more than a half-century of
beautiful experimental and theoretical work. At present 
their study is mainly aimed to inquire about possible exotic
properties of the neutrino associated  with its oscillations and
massiveness by means of exclusive and inclusive scattering
processes on nuclei, which are often used as neutrino detectors.
An example is given by the recent experiments performed by both
the LSND \cite{Ath98,Agu01}  and the KARMEN \cite{Eit99}
Collaborations, looking for $\nu_\mu\rightarrow\nu_e$ and
$\tilde{\nu}_\mu\rightarrow
\tilde{\nu}_e$ oscillations with neutrinos produced by
accelerators.
When the $\nu_e$ come from decay-at-rest (DAR) of $\mu^+$
the  flux contains neutrinos with a maximum energy energy of $50$
MeV and can be detected through both  exclusive
 and  inclusive $\nu_e+ {^{12}C} \rightarrow
{^{12}}N+ e^-$  reactions \cite{Aue01}. In the case of
$\nu_\mu$  coming from the decay-in-flight (DIF) of $\pi^+$,
the neutrino flux extends over the range $(0,300)$ MeV,
and the $\nu_\mu \rightarrow \nu_e$ appearance mode is looked for
experimentally through the reaction $\nu_{\mu}+ {^{12}C} \rightarrow
{^{12}N}+ \mu^-$, which also has been measured \cite{Aue02a}.
On the other hand we don't have at our disposal experimental
information on the
$\nu_e$-reaction in the DIF energy range,
which is necessary in the evaluation of the  oscillation
probabilities. Therefore, it is imperative
to develop nuclear models, capable of reproducing the
$\nu_e$-DAR and
$\nu_\mu$-DIF data, to be used
to predict reliable values for the $\nu_e+ {^{12}C} \rightarrow
{^{12}N}+ e^-$ cross section in the DIF energy range, and  to
 calibrate forthwith the
$\nu_\mu \rightarrow \nu_e$ appearance probability. Needless to
say, in addition and for consistence, the implemented model
should also describe properly the well-known
${^{12}N} \rightarrow {^{12}C}+ \nu_e+ e^+$
$\beta^+$-decay and the $\mu^-+{^{12}C}\rightarrow \nu_{\mu}+{^{12}B}$
muon-capture modes.

The weak interaction formalism most frequently employed in the literature
is that of Donnelly  and Walecka \cite{Don79,Wal95},
where  the  nuclear form factors  are classified as
Coulomb ($\cal M$), longitudinal ($\cal L$),
transverse electric (${\cal T}^{el}$) and transverse magnetic (${\cal
T}^{mag}$),  in close analogy with the
electromagnetic transitions.
They in turn depend on seven nuclear matrix elements, denoted as:
$M_{J}^{M}, \Delta_{J}^{{M}}, {\Delta'}_{J}^{{M}},
\Sigma_{J}^{{M}}, {\Sigma'}_{  J}^{{M}},
{\Sigma''}_{  J}^{   {M}}$ and $\Omega_{  J}^{  {M}}$.
However, in  studying neutrino induced reactions
\cite{Aue97,Vol00} it is sometimes preferred to employ the
formulation done by Kuramoto  \etal \cite{Kur90}, mainly
because of its simplicity.
The later formalism  does not include  the velocity
dependent terms in the hadronic current, and therefore the reaction
cross section only
depends on  three nuclear form factors, denoted as 
$|\bra{f}{\tilde 1}\ket{i}|^2, |\bra{f}{\tilde\sigma}\ket{i}|^2$ 
and $\Lambda$. 
What is more, the formalism of Kuramoto 
 \etal \cite{Kur90} does not include the muon capture 
rates. Therefore, to describe
simultaneously the
neutrino-nucleus reactions and  $\mu$-capture processes
it is necessary to resort to additional theoretical developments, 
such as those of Luyten \etal \cite{Luy63} and  
Auerbach and Klein \cite{Aue83}, where one uses
the matrix elements $M_V^2, M_A^2$ and $M_P^2$ that
are related to the former ones in a non-trivial way.

Quite recently, we have carried  out a multipole expansion of
the $V-A$ hadronic current, similar   to the one used  by
Barbero {\it et al.} \cite{Bar98} for the neutrinoless double beta
decay,  expressing all above mentioned  observables in terms of
the vector ($V$) and axial vector ($A$)  nuclear
 form factors $\M_V$ and $\M_A^M$, with $M=-1,0,+1$ \cite{Krm02}.
 Such a  classification  is closely related
to the $V-A$ structure of the weak current and to the
currently used nuclear $\beta$-decay
formalisms, where  the nuclear moments are expressed in a way that the
{\it forbiddeness} of the transitions  is easily
recognized.
This in no way means that we have developed a new theoretical framework;
the final results can be found in one form or another in
the literature. The main difference
stems from the fact that we use the Racah algebra from the start,
employing the spherical spatial coordinates  instead of
the cartesian ones. Concomitantly, we also express the
lepton trace in spherical coordinates, as done for instance in
the book of Supek \cite{Sup64}.
Here we present a few more details than in Ref. \cite{Krm02} on the
procedure which we have followed
and also consider the first order nonlocal corrections,
which give rise to the additional matrix elements $\M_{V'}^M$ and
$\M_{A'}$.

In Ref. \cite{Krm02} we have analyzed the
inconveniences  that  appear in applying  RPA-like models 
to describe the nuclear structure of the
$\{{{^{12}B},{^{12}C},{^{12}N}}\}$ triad. We have  established that
the projected quasiparticle RPA (PQRPA)
was  the proper approach to treat  both the short range  pairing and
the long range RPA correlations. More details are given in the present work,
putting special emphasis on  the differences between  the
projected and the usual QRPA approximations.
We also compare our PQRPA  cross sections with  recent shell model
(SM) calculations \cite{Hay00,Aue02}, and analyze the reliability of
the calculated energy dependence of the
 $\nu_e+ {^{12}C} \rightarrow {^{12}N}+e^-$ cross-section in the DIF
energy region to be used in the evaluation of the oscillation probabilities.

The work will be organized as follows. In Section II we explain our
multipole decomposition of the weak current and present the
corresponding formula for the neutrino-nucleus cross section and the
 $\mu$-capture.
In Section III we briefly overview the projected BCS (PBCS) and the
PQRPA equations. The main reason for the last is to  emphasize the
relationship with the usual QRPA approximation, which is difficult to
find  in the existing literature.
Finally, in Section IV we show the numerical results which come from 
the PQRPA model and compare them
with the most recent shell model calculations.

\section{Formalism for the Weak Interacting Processes}
\label{sec2}
The weak Hamiltonian is expressed in the form \cite{Don79,Wal95,Bli66,Bar98}
\br
H_{{\sss {W}}}(\rb)&=&\frac{G}{\sq}J_\alpha l_\alpha e^{-i\rb\cdot\kb},
\label{1}\er
where $G=(3.04545\pm 0.00006){\times} 10^{-12}$ is the Fermi coupling
constant (in natural units),

\br
J_\alpha&=&i\g_4
\left[\gV\g_\alpha-\frac{\gM}{2\Mass}\s_{\alpha\beta}k_\beta
+\gA\g_\alpha\g_5+i\frac{\gP}{\mass_\ell} k_\alpha\g_5\right]\equiv
(\Jb,iJ_\emptyset)\label{2}\er
is the hadronic current operator
\footnote{To avoid confusion, we will be using roman fonts
($\Mass$,$\mass$) for masses and math italic fonts ($M$,$m$) for
azimuthal quantum numbers.},  and
\br
l_\alpha&=&-i\overline{u}_{s_\ell}(\pb,E_\ell)\g_\alpha(1+\g_5)u_{s_\nu}
(\qb,E_\nu)\equiv
(\lb,il_\emptyset)
\label{3}\er
is the plane wave approximation for the matrix element of the
leptonic current in the case
of neutrino reactions. Here $\alpha, \beta = 1,2,3,4$, and 
Walecka's notation \cite{Wal95} with the Euclidean
metric for quadrivectors is
employed, \ie  $x= \{ \xb,x_4=i x_\emptyset \}$. The only difference
is that we substitute his indices
$(0,3)$ by our indices $(\emptyset,0)$, which means that
we use the index $\emptyset$ for the temporal component and 
the index $0$ for the third spherical component. The quantity
\br
k = P_i-P_f\equiv \{\kb,ik_\emptyset \}
\label{4}\er
is the momentum transfer ($P_i$ and $P_f$ 
are momenta of the initial and final nucleon
(nucleus)), ${\rm M}$ is the nucleon mass, ${\rm m}_\ell$ is 
the mass of the charged lepton,
and $g_{\sss V}$, $g_{\sss A}$, $g_{\sss M}$ 
and $g_{\sss P}$ are, respectively, the vector,
axial-vector, weak-magnetism and pseudoscalar effective 
dimensionless coupling constants.
Their numerical values are 
\br
g_{\sss V}&=&1 ;
~g_{\sss A}=1.26 ; \nn\\
g_{\sss M}&=&\kappa_p-\kappa_n=3.70 ;
~g_{\sss P}=g_{\sss A}\frac{2\Mass \mass_\ell }{k^{2}+\mass_\pi^2}.
\label{5}
\er
These estimates for $g_{\sss M}$ and $g_{\sss P}$ come from the conserved
vector current (CVC) hypothesis, and from the partially conserved axial vector current
(PCAC) hypothesis,  respectively. In the numerical calculation we will use an effective
axial-vector coupling
$g_{\sss A}^{\sss}=1$ \cite{Bro85,Cas87,Ost92}.

The finite nuclear size (FNS) effect is incorporated via the dipole form factor
with a cutoff $\Lambda=850$ MeV, \ie as:
\[
g\go g\left( \frac{\Lambda^{2}}{\Lambda^{2}+k^{2}}\right)^{2}.
\]

To use \rf{1} with  the non-relativistic nuclear wave
functions, the Foldy-Wouthuysen transformation has to be performed
on the hadronic current \rf{2}. When the velocity dependent
terms
are included
 this yields:
\br
J_\emptyset&=&\gV + (\ga+\gpa) {\mbs} \cdot\hat{\kb}
- \gA \mbs \cdot \vb,
\nonumber\\
\Jb&=&
-\gA {\mbs} -i\gw {\mbs}\x\hat{\kb}-\gv \hat{\kb}
+\gpb({\mbs} \cdot\hat{\kb})\hat{\kb}+{\gV}
\vb ,\nn\\
\label{6}\er
where $\vb\equiv -i \mbn/\Mass$ is the velocity operator, acting  on the nuclear wave
functions.
The following short notation has been introduced:
\br
\gv&=&\gV\frac{\k}{2\Mass};~
\ga=\gA\frac{\k}{2\Mass};~
\gw=(\gV+\gM)\frac{\k}{2\Mass},
\nn\\
\gpa&=&\gP\frac{\k}{2\Mass}\frac{k_\emptyset}{\mass_\ell};~
\gpb=\gP\frac{\k}{2\Mass}\frac{\k}{\mass_\ell},
\label{7}\er
where $\k\equiv |\kb|$.

 In performing the multipole expansion of the nuclear operator
\be
O_\alpha\equiv (\Ob,iO_\emptyset)=J_\alpha e^{-i\kr},
\label{8}\ee
the momentum $\kb$ is taken to be along  the $z$ axis, \ie
\br
e^{-i\kr}&=&\sum_{L}i^{-L}\sqrt{4\pi(2L+1)}j_L(\k r)
Y_{L0}(\hat{\rb}),
\label{9}\er
and one gets
\br
O_\emptyset&=&\sum_{J}i^{-J}\sqrt{4\pi(2J+1)}j_J(\k r)Y_{J0}(\hat{\rb})
J_\emptyset,
\nn\\
O_M&=&\sum_{JL}i^{-L}F_{JLM}\sqrt{4\pi(2J+1)}j_L(\k r)
\left[Y_{L}(\hat{\rb})\otimes{\Jb}\right]_{JM}.
\label{10}\er
The geometrical factors
\br
F_{JLM}&\equiv&(-) ^{J+M}\sqrt{(2L+1)}
\threej{L}{1}{J}{0}{-M}{M},
\label{11}\er
which will be seldom used in our work,  fulfill the sum rule
 \br
&&\sum_{L}F_{JLM}F_{JLM'}=\del_{MM'},
\label{12}\er
and  their explicit values are listed in Table I.

\begin {table}[h]
\caption {Values of the geometrical factors $F_{JLM}$.}
\newcommand{\cc}[1]{\multicolumn{1}{c}{#1}}
\renewcommand{\tabcolsep}{2.pc} 
\renewcommand{\arraystretch}{1.2} 
\label{table1}
\bigskip
\begin{center}
\begin{tabular}{|c|c|c|}
\hline
 $M$  & $L$  & $F_{JLM}$\\
\hline\hline
$ 0$  &$J+1$ & $-\sqrt{\frac{J+1}{2J+1}}$ \\ \hline
$ 0$  &$J$   & $0$ \\ \hline
$ 0$  &$J-1$ & $\sqrt{\frac{J}{2J+1}}$ \\ \hline
$ 1$  &$J+1$ & $\sqi\sqrt{\frac{J}{2J+1}}$ \\ \hline
$ 1$  &$J$   & $-\sqi$ \\ \hline
$ 1$  &$J-1$ & $\sqi\sqrt{\frac{J+1}{2J+1}}$ \\ \hline
$-1$  &$J+1$ & $\sqi\sqrt{\frac{J}{2J+1}}$ \\ \hline
$-1$  &$J$   & $\sqi$ \\ \hline
$-1$  &$J-1$ & $\sqi\sqrt{\frac{J+1}{2J+1}}$ \\ \hline
\end{tabular}
\end{center}\end {table}

After some Racah algebra we find
\be
O_\alpha=\sum_{J}\sqrt{2J+1} {\sf O}_\alpha(J), 
\label{13}\ee
with
\br
 {\sf O}_\emptyset(J)&=&i^{-J} \sqrt{4\pi}
\left[
\gV{\sf Y}_{J0}(\k\rb)-\gA {\sf Y}_{J0}(\k\rb,\mbs\cdot\vb)\right]
+\sqrt{4\pi}\left(\ga +\gpa \right)\sum_{L=J\pm1}i^{-L}F_{JL{0}}{\sf S}_{JL{0}}(\k\rb)
\nn\\
\label{14}\er
and
\br
 {\sf O}_{M}(J) &=&\sqrt{4\pi}
\sum_{L}i ^{-L} F_{JLM}\left[
-\gA {\sf S}_{JLM}(\k\rb)+\gV {\sf P}_{JLM}(\k\rb)
-i(-)^J\gv F_{JL0} {{\sf Y}_{JM}(\k\rb)}\right.
\nn\\
&&\hspace{3.2cm}+\sum_{I}
\left.\left(i(-)^L\gw G_{JLI}+\gpb F_{JL{0}}F_{JI{0}}\right)
{\sf S}_{JIM }(\k\rb){\over}
\right],
\label{15}\er
where we have introduced the operators
listed in Table \ref{TableII},
\begin{table}[t]
\caption{Elemental operators and their parities.\label{TableII}}
\newcommand{\cc}[1]{\multicolumn{1}{c}{#1}}
\renewcommand{\tabcolsep}{2.0pc} 
\renewcommand{\arraystretch}{1.2} 
\bigskip
\begin{center}
\begin{tabular}{|c|c|}
\hline
Operator&Parity
\\
\hline \hline ${\sf Y}_{JM}(\k\rb)=j_J(\k r)
Y_{JM}(\hat{\rb})$&$(-)^{J}$
\\ \hline
$ {\sf S}_{JLM}(\k\rb)= j_L(\k r)
 \left[Y_{L}(\hat{\rb})\otimes{\mbs}\right]_{JM}
$&$(-)^{L}$ 
\\ \hline 
${\sf P}_{JLM}(\k\rb)=j_L(\k r)
[ Y_L(\hat{\rb})\otimes\vb ]_{JM}
$ &$(-)^{L+1}$
\\ \hline
$ {\sf Y}_{JM}(\k\rb,\mbs\cdot\vb)=
j_J(\k r) Y_{JM}(\hat{\rb})(\mbs \cdot \vb)
$&$(-)^{J+1}$
\\ \hline
\end{tabular}
\end{center}
\end{table}
and  the coefficients
\br
G_{JLI}&=&(-)^{J}
\sqrt{6(2L+1)(2I+1)}{\sixj{1}{1}{1}{I}{L}{J}}
\threej{I}{1}{L}{0}{0}{0},
\label{16}\er
which obey the sum rule
\br
&&\sum_{L}F_{JLM}G_{JLI}
=-M F_{JIM}.
\label{17}\er

Using \rf{12} and \rf{17} we can  rewrite the spherical components of
${\sf O}_{M} (J)$ as
\br
{\sf O}_{M} (J)&=&\sqrt{4\pi}
\sum_{L}i^{-L} F_{JLM}\left [{\over}
(-\gA+M \gw+\gpb\del_{M0})  {\sf S}_{JLM}(\k\rb)+\gV {\sf P}_{JLM}(\k\rb)
\right]\nn\\
&-& \sqrt{4\pi}i^{-J} \gv \del_{M0} {{\sf Y}_{J0}(\k\rb)}.
\label{18}\er

For the neutrino-nucleus reaction the momentum transfer is
$k=p_\ell-q_\nu$, with
 $p_\ell\equiv\{\pb_\ell,iE_\ell\}$
and $q_\nu\equiv\{\qb_\nu,iE_{\nu}\}$, and the corresponding cross
section reads 
\br
\s(E_\ell,J_f)& = &\frac{|\pb_\ell| E_\ell}{2\pi} F(Z+1,E_\ell)
\int_{-1}^1
d(\cos\theta)\T_{\s}(\k,J_f),\nn\\
\label{19}\er
where $F(Z+1,E_\ell)$ is  the Fermi function,
$\theta\equiv \hat{\qb}_\nu\cdot\hat{\pb}_\ell$, and
\br
\T_{\s}(\k,J_f)\equiv
\frac{1}{2J_i+1}
\sum_{ s_\ell,s_\nu }\sum_{M_i,  M_f }
\left|\bra{J_fM_f}H_{{\sss {W}}}\ket{J_iM_i}\right|^{2},
\label{20}\er
with  $\ket{J_iM_i}$ and $\ket{J_fM_f}$
being the nuclear initial and final state vectors.

The weak hamiltonian matrix element reads
\br
\bra{J_fM_f}H_{{\sss {W}}}\ket{J_iM_i}
&=&\frac{G}{\sq}{\cal O}_\alpha l_\alpha,
\label{21}\er
where
\br
{\cal O}_\alpha\equiv \bra{J_fM_f}{ O}_\a\ket{J_iM_i},
\label{22}\er
and $l_\alpha$ is the leptonic current defined in \rf{3}.
Thus
\br
\T_{\s}(\k,J_f)&=&
\frac{G ^{2}}{2J_i+1}\sum_{M_i M_f}
{\cal O}_\alpha{\cal O}_\beta^{*}\L_{\alpha\beta},
\label{23}\er
where, the lepton trace  $\L_{\alpha\beta}$,
when expressed  in cartesian spatial coordinates, reads
\br
\L_{\alpha\beta}=
\frac{1}{2}
\sum_{s_\ell s_\nu }l_\alpha l^*_{ \beta}
&=&-\frac{1}{E_\ell E_\nu }\left[p_\a q_\b+q _\a p_\b
-\del_{\a\b}(p\cdot q)\pm \e_{\a\b\g\del} q_\g p_\del \right],
\label{24}\er
the positive (negative) sign standing for neutrino (antineutrino) scattering.

It is convenient to follow Ref. \cite{Sup64} and  express the spatial parts of
${\cal O}$ and $\L$ in spherical coordinates
($M, M'=0, -1, 1$). In this way one might write
\br
\T_{\s}(\k,J_f)&=&
\frac{G^2}{2J_i+1}\sum_{M_fM_i}\left[
\left| {\cal O}_\emptyset\right|^{2}\L_{\emptyset\emptyset}
+\sum_{MM'}{\cal O}_M {\cal O}_{M'}^{*}\L_{MM'}
-2\Re\left( {\cal O}_\emptyset^{*}\sum_{M}(-)^{M}{\cal O}_{-M}
 \L_{\emptyset M}\right)\right],
\nn\\
\label{25}\er
with  \cite{Sup64}
\br
\L_{\emptyset\emptyset}\equiv \L_{44}&=&1+\frac{\pb\cdot \qb
}{E_\ell E_\nu },
\label{26}\er
\br
\L_{\emptyset M}\equiv -i \L_{4 M}&=&\frac{1}{E_\ell
E_\nu }\left[E_\ell q_M+E_\nu  p_M \mp i(\qb\x\pb)_M\right],
\label{27}\er
\br
\L_{MM'}&=&\del_{MM'}+\frac{1}{E_\ell E_\nu }
\left(q^*_{M}p_{M'}+q_{M'}p_{M}^*
-\del_{MM'}\pb \cdot\qb    \right)
\nn\\
&\pm&\sqrt{6}(-)^{M}\threej{1}{1}{1}{-M}{M'}{M-M'}
\left(\frac{q_{M-M'}^*} {E_\nu}-\frac{p_{M-M'}^*}{E_\ell}\right).
\label{28}\er

From the Wigner-Eckart theorem we also get:
\br
{\cal O}_{\emptyset}&=&\sum_{J}(-)^{J_f-M_f}\sqrt{2J+1}\threej{J_f}{J}{J_i}{-M_f}{0}{M_i}
\Bra{J_f}{\sf O}_\emptyset(J)\Ket{J_i},
\label{29}\er
and
\br
{\cal O}_{M}&=&\sum_{J}(-)^{J_f-M_f}\sqrt{2J+1}
\threej{J_f}{J}{J_i}{-M_f}{M}{M_i}
\Bra{J_f}{\sf O}_M(J)\Ket{J_i}.
\label{30}\er
Using now the orthogonality condition
\br
\sum_{M_iM_f}\threej{J_f}{J}{J_i}{-M_f}{M}{M_i}
\threej{J_f}{J'}{J_i}{-M_f}{M'}{M_i}&=&\frac{1}{(2J+1)}\del_{JJ'}\del_{MM'}
\label{31}\er
one obtains
\br
\T_{\s}(\k,J_f)&=&
\frac{G^2}{2J_i+1}\sum_{J}\left[
|\Bra{J_f}{\sf O}_\emptyset(J)\Ket{J_i}|^2\L_{\emptyset\emptyset}
+\sum_{M}|\Bra{J_f}{\sf O}_M(J)\Ket{J_i}|^{2}\L_{MM}
\right.
\nn \\
&&\hspace{2cm}
\left.-2\Re\left(\Bra{J_f}{\sf O}_\emptyset(J)\Ket{J_i}^{*}\Bra{J_f}{\sf O}_0(J)\Ket{J_i}
\L_{\emptyset0}\right) \right],
\label{32}\er
where, from \rf{14} and \rf{18},
\br
\Bra{J_f}{\sf O}_\emptyset(J)\Ket{J_i}&=&\sqrt{2J_i+1}
\left[\gV\M_{\sss V}(J)-\gA \M_{\sss A'}(J)+\left(\ga +\gpa \right)
\M_{\sss A}^0(J)
\right]
\nn\\
\Bra{J_f}{\sf O}_{M} (J)\Ket{J_i}&=&\sqrt{2J_i+1}
\left [(-\gA+M \gw+\gpb\del_{M0}) \M_{\sss A}^M(J)
+\gV \M_{\sss V'}^{M}(J)-\gv \del_{M0} \M_{\sss V}(J)\right],
\nn\\
\label{33}\er
with the nuclear matrix elements defined as
\br
\M_{\sss V}(J)&=&i^{-J}\sqrt{\frac{4\pi}{2J_i+1}}
\Bra{J_f}{\sf Y}_{J}(\k\rb)\Ket{J_i},
\nn\\
\M_{\sss A}^{M}(J)&=&\sqrt{\frac{4\pi}{2J_i+1}}\sum_{L}i^{-L}
F_{JL{M}}\Bra{J_f}{\sf S}_{JL}(\k\rb)\Ket{J_i},
\label{34}\\
\M_{\sss A'}(J)&=&i^{-J}\sqrt{\frac{4\pi}{2J_i+1}}
\Bra{J_f}{\sf Y}_{J}(\k\rb,\mbs\cdot\vb)\Ket{J_i},
\nn\\
\M_{\sss V'}^{M}(J)&=&\sqrt{\frac{4\pi}{2J_i+1}}\sum_{L}i^{-L}
 F_{JL{M}}\Bra{J_f}{\sf P}_{JL}(\k\rb)\Ket{J_i}.
\nn\er
The explicit expressions for  $\L_{\emptyset M}$ and $\L_{MM'}$ that
appear in \rf{32} are \cite{Krm02}:
\br
\L_{\emptyset 0}&=&\left(\frac{q_0}{E_\nu}+\frac{p_0}{E_\ell}\right),
\nn\\
\L_{00}&=&1+\frac{2q_0p_0-\pb\cdot\qb}{E_\ell E_\nu},
\nn\\
\L_{\pm1,\pm1}&=&1-\frac{q_0p_0}{E_\ell E_\nu}\pm
\left(\frac{q_0}{E_\nu}-\frac{p_0}{E_\ell}\right),
\label{35}\er
with
\br
q_0&=&{\kh}\cdot \qb=\frac{E_\nu(|\pb|\cos\theta-E_\nu)}{\k},
\nn\\
p_0&=&{\kh}\cdot \pb=\frac{|\pb|(|\pb|-E_\nu\cos\theta)}{\k}.
\label{36}\er
Finally, the transition amplitude  can be cast in the form:
\br
&&\T_{\s}(\k,J_f)={G^2}
\sum_J\left\{ \L_{\emptyset\emptyset}
 \left[g_{{\sss {V}}}^2 |\M_{\sss V} (J)|^2
+\left|\left(\overline{g} _{{\sss {A}}}+\overline{g} _{{\sss {P1}}}\right)
\M_{\sss A}^0(J)-
g_{{\sss {A}}} \M_{\sss A'}(J)\right|^2
\right]\right.
\nn\\
&&+\L_{00} \left[ \Re\left[\left(\gv \M_{{\sss {V}}}(J)
-2\gV \M_{{\sss {V'}}}^0(J)\right)\gv\M^*_{{\sss {V}}}(J)\right]
+\left( \gpb^2-2\gA\gpb\right)|\M_{{\sss {A}}}^0 (J)|^2 \right]
\nn\\&&+\sum_{{M}=0,\pm1} \L_{MM}
\left|{\over}\left(\gA - M \gw\right)\M_{\sss A}^{M}(J)-\gV
\M_{\sss {V'}}^{M}(J){\over}\right|^2
\nn\\&&+2\L_{\emptyset 0} \Re\left(\gV\left[{\over}\gv\M_{\sss V}(J)
- g_{{\sss V}}
\M_{\sss {V'}}^0(J){\over}\right]\M_{\sss V}^*(J)\right.
\nn\\&&\hspace{1.4cm}+\left(\gA-\gpb\right)
\left[(\ga+\gpa)\M_{\sss A}^0(J)-\left.\left.
\gA\M_{\sss A'}(J){\over}\right]\M_{\sss A}^{0*} (J){\over}\right){\over} \right\}.
\label{37}\er

The muon capture transition amplitude $\T_{\Lambda}(J_f)$ can be
derived from the result \rf{32} for
the neutrino-nucleus reaction amplitude, by keeping  in mind
that:
 i) the roles of $p$ and $q$ are interchanged within  the matrix element of
the leptonic current, which brings in a minus sign in the last term of
$\L_{\pm1, \pm1}$,
ii) the momentum transfer turns out to be $k=q-p$, and therefore
the signs of the right-hand sides
of \rf{36}  have to be changed,
and iii) the threshold values ($\pb\go 0: \kb\go
\qb, k_\emptyset\go E_\nu-\mass_\ell$) must be used for the lepton traces. All this yields:
\be
\L_{\emptyset\emptyset}=\L_{00}=\L_{\emptyset0}=1
,~~\L_{\pm 1,\pm 1}= 1\mp 1.
\label{38}\ee
One should also remember that instead of summing
over the initial lepton spins $s_\ell$, as done in \rf{20}, one
has now to average over the same quantum number, getting
\br
\Lambda(J_f)&=&\frac{E_\nu^2}{2\pi}|\phi_{1S}|^2\T_{\Lambda}(J_f),
\label{39}\er
where $\phi_{1S}$ is the  muonic bound state wave
function evaluated at the origin, and
$E_\nu=\mass_\mu-(\Mass_n-\Mass_p)-E_B^\mu-E_f+E_i$, where $E_B^\mu$ is the
binding energy of the muon in the $1S$ orbit. In view of \rf{39} the
transition amplitude reads:
\br
\T_{\Lambda}(J_f)
&=&
\frac{G^2}{2J_i+1}\sum_{J}\left[
|\Bra{J_f}{\sf O}_\emptyset(J)-{\sf O}_0(J)\Ket{J_i}|^2
+2|\Bra{J_f}{\sf O}_{-1}(J)\Ket{J_i}|^{2}
\right],
\label{40}\er
which, based on the parity considerations, can be expressed as
\br
\T_{\Lambda}(J_f)
&=&G^2\sum_J \left\{
\left|(\gV+\gv) \M_{\sss V}(J)-\gV
\M_{\sss{V'}}^0(J)\right|^2 \right.
+\left|(\gA+\ga-\gp) \M_{\sss A}^0(J)-\gA
\M_{\sss A'}(J)\right|^2
\nn\\
&&\hspace{1.3 cm}+\left.2\left|(\gA+\gw) \M_{\sss A}^{-1}(J)-\gV
\M_{\sss {V'}}^{-1}(J)\right|^2\right\},
\label{41}\er
where $\gp=\gpb-\gpa$.
In the case of muon capture, it is convenient to rewrite
the effective coupling constants as
\cite{Luy63}
\br
\gv&=&\gV\frac{E_\nu}{2\Mass};~\ga=\gA\frac{E_\nu}{2\Mass};\nn\\
\gw&=&(\gV+\gM)\frac{E_\nu}{2\Mass};~ \gpa=\gP\frac{E_\nu}{2\Mass}.
\label{42}\er

Lastly, we mention that the $B$-values for 
the Gamow-Teller (GT) beta transitions are
defined and related to the ft-values as \cite{Tow95}:
\br
\frac{|g_{\sss A}\Bra{J_f}\s\Ket{J_i}|^{2}}{2J_i+1}\equiv
B(GT)&=&\frac{6146}{ft}~\mbox{sec}.
\label{43}\er

Let us now compare our matrix elements with those 
currently used in the literature. First, in 
the Walecka's notation (see Eqs. (45.13) in
Ref. \cite{Wal95})
one has
\br
 {\sf O}_\emptyset(J)&=&{\cal M}_{J0},
\nn\\
 {\sf O}_{M}(J)&=&
 \left\{
\begin{array}{ccc}
{\cal L}_{J0},\;\;& \mbox{for} \;& M=0 \\
-\sqi\left[M{\cal T}^{mag}_{JM}
+{\cal T}^{el}_{JM}\right],\;\;& \mbox{for}\;& M=\pm 1
\end{array}\right.,
\label{44}\er
where the meaning of ${\cal M}_{J0}$ and ${\cal L}_{J0}$ is self evident, while
\br
{\cal T}^{el}_{JM}  &=&
\gw{\sf S}_{JJM}-
i^J\sq \sum_{L=J\pm 1}i^{-L}F_{JLM}\left [{\over}
\gA {\sf S}_{JLM}-\gV {\sf P}_{JLM}
\right],
\nn\\
{\cal T}^{mag}_{JM}  &=&
-\gA {\sf S}_{JJM}+\gV {\sf P}_{JJM}
+i^J\sq \gw\sum_{L=J\pm 1}i^{-L}F_{JLM}{\sf S}_{JLM}.
\label{45}\er
The matrix elements defined by Donelly  \cite{Don79} are related to
ours as:
\br
M_{J}^{M}(\k \rb)
&=& {\sf Y}_{JM}(\k\rb),
\nn \\
\Delta_{J}^{{M}}(\k \rb)
&=&
\left(\frac{i \Mass}{\k}\right) {\sf P}_{JJM}(\k\rb),
\nn\\
{\Delta'}_{J}^{{M}}(\k \rb)
&=&i^J\sq
\left(\frac{\Mass}{\k}\right)
\sum_{L=J\pm1}i^{-L} F_{JLM} {\sf P}_{  {LJM}}(\k\rb),
\label{46}\\
\Sigma_{J}^{{M}}(\k \rb)
&=&{\sf S}_{JJM}(\k\rb),
\nn \\
{\Sigma'}_{  J}^{   {M}}(\k \rb)
&=& i^{J-1}{\sqrt{2}}
\sum_{ { L=J\pm1}}i^{-L} F_{JLM}{\sf S}_{  {JL{M}}}(\k\rb),
\nn \\
{\Sigma''}_{  J}^{   {M}}(\k \rb)
&=& i^{J-1}
\sum_{ { L=J\pm1}}i^{-L}  F_{JL0}{\sf S}_{  JL{M}}(\k\rb),
\nn\\
\Omega_{  J}^{  {M}}(\k \rb)
&=&
 \left(\frac{i\Mass}{\k} \right)
{\sf Y}_{  JM}(\k\rb, \mbs \cdot \vb).
\nn\er
The relationship between our formalism and those from Refs.
\cite{Kur90,Luy63} can be obtained from
the formula
\br
\threej{L}{1}{J}{0}{M}{-M}\threej{L'}{1}{J}{0}{M}{-M}&=&
\frac{\del_{LL'}}{3(2L+1)}
-M\sqrt{\frac{3}{2}}\threej{L}{L'}{1}{0}{0}{0}
\sixj{L}{1}{J}{1}{L'}{1}
\nn\\
&+&(-) ^{J+1}\sqrt{\frac{5}{6}}(3M^{2}-2)\threej{L}{L'}{2}{0}{0}{0}
\sixj{L}{1}{J}{1}{L'}{2}.
\nn\\
\label{47}\er
For the matrix elements of  Kuramoto \etal
\cite{Kur90} we get
\br
|\bra{f}\tilde{1}\ket{i}|^{2}
&=&\sum_{J}\left|\M_{\sss V}(J)\right|^2,
\nn\\
\bra{f}\tilde{\s}\ket{i}|^{2}
&=&\sum_J \sum_{M=0,\pm1}|\M_{\sss A}^M(J)|^2,
\nn\\
\Lambda
&=&\frac{1}{3}\sum_J\left(
|\M_{{\sss {A}}}^0(J)|^2
-|\M_{{\sss {A}}}^{1}(J)|^2\right);
\label{48}\er
and for those of Luyten \etal \cite{Luy63}
\br
M_V^2&=&\left(\frac{E_\nu}{m_\mu}\right)^2
\sum_{J}\left|\M_{\sss V}(J)\right|^2,
\nn\\
M_A^2&=&\left(\frac{E_\nu}{m_\mu}\right)^2
\sum_J \sum_{M=0,\pm1}|\M_{\sss A}^M(J)|^2,
\nn\\
M_P^2&=&\left(\frac{E_\nu}{m_\mu}\right)^2
\sum_J |\M_{\sss A}^0(J)|^2.
\label{49}\er

\section{Projected QRPA Formalism}
\label{sec3}

We have shown in  Ref. \cite{Krm02}  that  to account for the weak decay
observables  in  a light $N=Z$ nucleus in the framework of the RPA,
besides including
 the BCS correlations, it is imperative to perform the particle
number projection. 
It should be remembered that in heavy nuclei
the neutron excess is usually large, which makes 
the projection procedure less important that in 
light nuclei \cite{Krm93}. 
In this section we give a more detailed  overview of the
projected BCS (PBCS) and projected QRPA (PQRPA) approximations.

When the the excited states $\ket{J_f}$
in the final $(Z\pm 1,N\mp 1)$ nuclei
are described within the PQRPA, the transition amplitudes
for the multipole charge-exchange operators ${\sf Y}_{J}$, \etc,
listed in Table \ref{TableII}
, read
\br
\Bra{J_f, Z+\mu,N-\mu}{\sf Y}_{J}\Ket{0^+} & = & {1 \over
(I^{Z}I^N)^{1/2}}
\sum_{pn}
 \left[ {\Lambda_\mu(pnJ) \over
 (I^{Z-1+\mu}(p)I^{N-1+\mu}(n))^{1/2}}
 X_{\mu}^{\ast}(pnJ_f)\right.\nonumber\\
 &+&\left.{\Lambda_{-\mu}(pnJ) \over
 (I^{Z-1-\mu}(p)I^{N-1-\mu}(n))^{1/2}} Y_\mu^{\ast}(pnJ_f)\right],
\label{3.1} \er
with the one-body matrix elements given by
\begin{eqnarray}
&&\Lambda_\mu(pnJ)=-\frac{\Bra{p}{\sf Y}_{J}\Ket{n}}{\sqrt{2J+1}}
 \left\{
\begin{array}{l}
\up \vn,\;\; \mbox{for}\; \mu = +1 \\ \un \vp,\;\; \mbox{for}\; \mu
= -1 \
\\\end{array}\right.,
\label{3.2}
\end{eqnarray}
where
\br
I^{K}(k_1k_2\cdot\cdot k_n)&=& \frac{1}{2\pi i}
\oint \frac{dz}{z^{K+1}} \s_{k_1}\cdots \s_{k_n}
 \prod_{k}(u_k^2 + z^2 v_k^2)^{j_k+1/2};~\s_{k}^{-1}=u^2_k+z^2_kv_k^2,
\label{3.3}\end{eqnarray}
are the  PBCS number projection integrals.

The PBCS gap equations are

\br 2\bar{e}_k u_k v_k-\Delta_k(u^2_k-v^2_k)&=&0,
\label{3.4}\er
where
\begin{eqnarray}
&&\Delta_k=-\frac{1}{2}\sum_{k'}{(2j_{k'}+1)^{1/2}\over
(2j_k+1)^{1/2}}
u_{k'}v_{k'}
{\rm G}(kkk'k';0)\frac{I^{Z-2}(kk')}{I^Z}
\label{3.5}\er
are the pairing gaps, and
\br
{\bar e}_k &=& e_k \frac{I^{Z-2}(k)}{I^Z} +
\sum_{k'} {(2j_{k'}+1)^{1/2}\over (2j_k+1)^{1/2}} v_{k'}^2 {\rm
F}(kkk'k';0)\frac{I^{Z-4}(kk')} {I^Z}+ \Delta
e_k,\label{3.6}\er
are  the dressed  single-particle energies. The PBCS  
correction term $\Delta e_k$ can be found in Ref. \cite{Krm93},
and ${\rm F}$ and ${\rm G}$ stand for the
usual particle-hole (ph) and particle-particle (pp) matrix
elements, respectively.

The forward, $X_{\mu}$, and backward, $Y_{\mu}$,  PQRPA amplitudes
are obtained by solving the RPA equations

\begin{eqnarray}
\left(\begin{array}{cc} A_\mu &  B \\  -B^{*}&
 -A^{*}_{-\mu} \end{array}\right) \left(\begin{array}{l}
X_\mu \\ Y_\mu  \end{array}\right) = \omega_\mu
 \left(\begin{array}{l} X_\mu  \\
Y_\mu  \end{array}\right),
\label{3.7}\end{eqnarray}
with the PQRPA matrices defined as:
\br
&&A_\mu(pn,p'n';J)=(\varepsilon^{Z-1+\mu}_p+\varepsilon^{N-1-\mu}_n
)\del_{pn,p'n'}
+ N_\mu(pn)^{-1/2}N_\mu(p'n')^{-1/2}\nn\\
&\times&\left\{[u_pv_nu_{p'}v_{n'}I^{Z-1+\mu}(pp')I^{N-3+\mu}(nn')
+v_pu_nv_{p'}u_{n'}I^{Z-3+\mu}(pp')I^{N-1+\mu}(nn')]{\rm
F}(pn,p´n´;J)\right. \nn\\
&+&\left.[u_pu_nu_{p'}u_{n'}I^{Z-1+\mu}(pp')I^{N-1+\mu}(nn')
+v_pv_nv_{p'}v_{n'}I^{Z-3+\mu}(pp')I^{N-3+\mu}(nn')]{\rm
G}(pn,p´n´;J)\right\}, \nn\\
&&B(pn,p'n';J)=N_\mu(pn)^{-1/2}N_{-\mu}(p'n')^{-1/2}I^{Z-2}(pp')I^{N-2}(nn')\nonumber\\
&\times& [(v_pu_nu_{p'}v_{n'}+u_pv_nv_{p'}u_{n'}) {\rm
F}(pn,p´n´;J) +(u_pu_nv_{p'}v_{n'}+v_pv_nu_{p'}u_{n'}){\rm
G}(pn,p´n´;J)],
\label{3.8}\er
where
\br
N_\mu(pn)=I^{Z-1+\mu}(p)I^{N-1+\mu}(n),
\label{3.9}\er
are the norms,
\br
\varepsilon^{K}_k=
{R_0^K(k)+R_{11}^K(kk)\over I^K(k)}-\frac{R_0^K}{I^K}
\label{3.10}\er
are the projected quasiparticle energies, and
the quantities $R^K$ are defined as \cite{Krm93}
\br
&&R_0^K(k)=\sum_{k_1} (2j_{k_1}+1)v_{k_1}^2e_{k_1}I^{K-2}(k{k_1})
+\frac{1}{4}\sum_{{k_1}{k_2}}(2j_{k_1}+1)^{1/2}(2j_{{k_2}}+1)^{1/2}
\nn\\
&\x&\left[v_{k_1}^2v_{{k_2}}^2{\rm F}({k_1}{k_1}{k_2}{k_2};0)I^{K-4}({k_1}{k_2}k)
+u_{k_1}v_{{k_1}}u_{{k_2}}v_{{k_2}}{\rm G}({k_1}{k_1}{k_2}{k_2};0)I^{K-2}({k_1}{k_2}k)\right],\nn\\
&&R_{11}^K(kk)=e_{k}[u_{k}^2I^{{k_1}}(kk)-v_{k}^2I^{K-2}(kk)]
+\sum_{{k_1}}{(2j_{k_1}+1)^{1/2}\over(2j_{k}+1)^{1/2}}
\nn\\
&\x&\left\{v_{k_1}^2{\rm F}({k_1}{k_1}kk;0)
[u_{k}^2I^{K-2}({k_1}kk)-v_{k}^2I^{K-4}({k_1}kk)]
-u_{k_1}v_{{k_1}}u_{k}v_{k}{\rm G}({k_1}{k_1}kk;0)I^{K-2}({k_1}kk)\right\}.
\nn\\
\label{3.11}\er

It is worth to note that the PQRPA formalism is valid not only for
the particle-hole charge-exchange excitations  $(Z\pm 1,N\mp 1)$, but
also
for the charge-exchange pairing-vibrations  $(Z\pm 1,N\pm 1)$. In the
later case one has simply to do the replacement $\mu\go -\mu$ in the
neutron sector.

The usual gap equations are obtained from
Eqs. \rf{3.4}-\rf{3.6} by:
\begin{enumerate}
\item  Making the replacement $e_{k}\rightarrow e_{k} -
\lambda_{\rm k}$, with $\lambda_{\rm k }$ being the chemical
potential, and taking the limit  $I^K \go 1$.
That is, the Eq. \rf{3.4} remains as it is, but instead 
of \rf{3.5} and \rf{3.6} one has now  
\begin{eqnarray}
&&\Delta_k=-\frac{1}{2}\sum_{k'}{(2j_{k'}+1)^{1/2}\over
(2j_k+1)^{1/2}}
u_{k'}v_{k'}
{\rm G}(kkk'k';0),
\label{3.12}\er
and
\br
{\bar e}_k &=& e_k-\l_{\rm k}+
\sum_{k'} {(2j_{k'}+1)^{1/2}\over (2j_k+1)^{1/2}} v_{k'}^2 {\rm
F}(kkk'k';0).\label{3.13}\er
\item 
Imposing  the subsidiary conditions
\br
Z = \sum_{j_p} (2j_p+1)^2 v_{j_p}^2,~~ N = \sum_{j_n} (2j_n+1)^2 v_{j_n}^2,
\label{3.14}\er
 as the number of particles is not any more
a good quantum number.
\end{enumerate}

Finally, the plain QRPA equations are recovered from \rf{3.7} and
\rf{3.8} by: i) dropping the index $\mu$ and taking 
the limit $I^K \go 1$, and ii) substituting 
the unperturbed  PBCS energies 
by the BCS energies relative to the Fermi level, \ie by
\be
E^{(\pm)}_k=\pm E_{k}+\l_{\rm k},
\label{3.15}\ee
where
${E}_{k}=({\bar e}_k^2+\Delta_k^2)^{1/2}$
are the usual BCS quasiparticle 
energies. In this way the the unperturbed energies
in \rf{3.8} are replaced  as
\footnote{We note that there are misprints in \cite[(23)]{Krm02}.}
\be
\varepsilon^{Z-1+\mu}_{j_p}+\varepsilon^{N-1-\mu}_{j_n}
\go E_{j_p}+E_{j_n}+\mu(\l_{\rm p}-\l_{\rm n}).
\label{3.16}\ee

\section{Numerical Results and Discussion}

In this section, our theoretical results within the PQRPA are confronted 
with the
experimental data  for  the ${\mu^-}+ ^{12}B   \rightarrow
^{12}C+ \nu_\mu$ muon capture rates, as well as for the neutrino 
 cross sections
involving the DAR reaction $\nu_e + ^{12}C \rightarrow
{^{12}N + e^-}$ and DIF reaction ${\nu_\mu + ^{12}C} \rightarrow {^{12}N} +
\mu^-$.  We also exhibit our
 predictions for  the  $\nu_e+ ^{12} C \rightarrow
^{12}N+ e^-$ differential cross section for  ${\nu_e}$-energies
 in the DIF energy range. At variance with our previous work, we
consider here also  the velocity dependent matrix elements
$\M_{\sss A'}(J)$ and
$\M_{\sss V'}^{M}(J)$, defined in \rf{34}.

\begin{table}[h]
\caption{\footnotesize BCS and PBCS results for neutrons. $E^{exp}_j$
stand for the experimental
energies used in the fitting procedure, and $e_j$ are the resulting 
s.p.e.
The underlined   quasiparticle energies
correspond to single-hole excitations (for $j_h=1s_{1/2},1p_{3/2}$)  and to
single-particle excitations (for $j_p=1p_{1/2},
1d_{5/2},2s_{1/2},1d_{3/2},1f_{7/2},2p_{3/2},2p_{1/2,1f_{5/2}}$).
The non-underlined energies are   mostly two hole-one particle and two
particle-one hole excitations. The fitted values of the pairing strengths
$v^{pair}_s$ in units of MeV-fm$^3$ are also displayed.}
\begin{center}
\label{TableIII}
\newcommand{\cc}[1]{\multicolumn{1}{c}{#1}}
\renewcommand{\tabcolsep}{1.1pc} 
\bigskip
\begin{tabular}{|c| r| rrr|rrr|}\hline
$$&$$&$$&$BCS$&$$&$$&$PBCS$&$$
\\
$Shell$&$E^{exp}_j$
&$E_j^{(+)}$&$E_j^{(-)}$&$e_j$&$ \varepsilon^{N}_j$
&$- \varepsilon^{N-2}_j$&$e_j$
\\ \hline
$1s_{1/2}$&&$11.34$&$\underline{-35.13}$&$-23.58$  &$
19.93$&$\underline{-34.99}$& $-22.37$
\\
$1p_{3/2}$&$-18.72$&$-5.07$&$\underline{-18.72}$&$-7.80$ &
$-1.28$&$\underline{-18.73}$& $-7.24$
\\
$1p_{1/2}$&$-4.94$&$\underline{-4.94}$&$-18.85$&$-2.07$
&$\underline{-4.95}$&$-22.33$&$-1.51$
\\
$1d_{5/2}$&$-1.09$&$\underline{-1.09}$&$-22.70$&$2.12$
&$\underline{-1.09}$&$-26.82$ & $2.16$
\\
$2s_{1/2}$&$-1.85$&$\underline{-1.86}$&$-21.93$&$2.70$
&$\underline{-1.85}$&$-25.98$ & $2.68$
\\
$1d_{3/2}$&$2.72$&$\underline{2.72}$&$-26.51$ &$6.24$
&$\underline{2.73}$&$-30.79$  & $6.26$
\\
$1f_{7/2}$&$5.81$&$\underline{5.82}$&$-29.61$ &$8.14$
&$\underline{5.83}$&$-33.61$  & $8.17$
\\
$2p_{3/2}$&$7.17$&$\underline{7.18}$&$-30.98$ &$11.49$
&$\underline{7.16}$&$-35.23$  & $11.47$
\\
$2p_{1/2}$&&$\underline{12.89}$&$-36.69$&$17.30$
&$\underline{12.89}$&$-41.01$ & $17.32$
\\
$1f_{5/2}$&&$\underline{16.72}$&$-40.52$&$19.18$
&$\underline{16.72}$&$-44.58$ & $19.21$
\\\hline
$v_{{s}}^{{pair}}$&&& &$23.16$& &&$23.92$
\\\hline
\end{tabular}
\end{center}
\end{table}
The calculations shown here were done in the same way as in the
previous work \cite{Krm02}. That is,
for the residual interaction we adopted  the delta force,
 \br
V &=&-4 \pi \left(v_sP_s+v_tP_t\right)
\delta(r),
\label{4.1} \er
 which  has been used extensively in the literature
\cite{Hir90a,Hir90b,Krm92} for describing the
single and double beta decays.
The configuration space includes
the single-particle orbitals with $nl=(1s,1p,1d,2s,1f,2p)$
for both protons and neutrons.
\begin{table}[h]
\caption{\footnotesize Same as Table \ref{TableIII} but for protons.}
\begin{center}
\label{TableIV}
\newcommand{\cc}[1]{\multicolumn{1}{c}{#1}}
\renewcommand{\tabcolsep}{1.1pc} 
\bigskip
\begin{tabular}{|c|c|ccc|ccc|} \hline
&&&$BCS$&&&$PBCS$&
\\
Shell&$E^{\rm exp}_j$&$E_j^{(+)}$&$E_j^{(-)}$&$e_j$
&$ \varepsilon^{Z}_j$
&$- \varepsilon^{Z-2}_j$&$e_j$
\\ \hline
$1s_{1/2}$&&$14.94$&$\underline{-33.13}$  &$-21.62$   &$19.44$&
$\underline{-32.99}$  &$-20.41$
\\
$1p_{3/2}$&$-15.95$ &$-2.24$&$\underline{-15.96}$  &$-4.95$    &$1.46$&$\underline{-15.95}$   &$-4.40$
\\
$1p_{1/2}$&$-1.94$  &$\underline{-1.95}$&$-16.25$  &$1.01$     &$\underline{-1.95}$&$-19.76$  &$1.56$
\\
$1d_{5/2}$&$1.61$   &$\underline{1.61}$&$-19.80$   &$4.76$     &$\underline{1.61}$&$-23.90$   &$4.83$
\\
$2s_{1/2}$&$0.42$   &$\underline{0.42}$&$-18.61$   &$4.88$     &$\underline{0.42}$&$-22.59$   &$4.88$
\\
$1d_{3/2}$&$4.95$   &$\underline{4.95}$&$-23.14$   &$8.40$     &$\underline{4.95}$&$-27.37$   &$8.43$
\\
$1f_{7/2}$&$8.42$   &$\underline{8.42}$&$-26.61$   &$10.72$    &$\underline{8.42}$&$-30.58$   &$10.75$
\\
$2p_{3/2}$&$9.76$   &$\underline{9.77}$&$-27.96$   &$14.05$    &$\underline{9.77}$&$-32.22$   &$14.07$
\\
$2p_{1/2}$&&$\underline{16.23}$&$-34.43$  &$20.63$    &$\underline{16.22}$&$-38.73$  &$20.66$
\\
$1f_{5/2}$&&$\underline{19.70}$&$-37.90$  &$22.15$    &$\underline{19.71}$&$-41.96$  &$22.20$
\\\hline
$v_{\sss{s}}^{\sss{pair}}$&&&&$23.13$&           &&$23.92 $
\\\hline
\end{tabular}
\end{center}
\end{table}

Most of the  bare single-particle energies (s.p.e.) $e_j$, as well as  
the value of  the singlet strength
within the pairing  channel ($v^{pair}_s$),
were fixed from the experimental
energies $E^{\rm exp}_j$ of the odd-mass nuclei  
$^{11}C$, $^{11}B$, $^{13}C$ and $^{13}N$.
That is, in the BCS case: 1) we  assume that the ground states in $^{11}C$
and $^{11}B$
are pure quasi-hole excitations ${E}^{(-)}_{j_h}$, with $j_h=1p_{3/2}$, and that
the lowest observed ${1/2}^{-}, {5/2}^{+},{1/2}^{+},{3/2}^{+}, {7/2}^{-}$ and
${3/2}^{-}$ states  in $^{13}C$ and $^{13}N$ are pure quasi-particle 
excitations
${E}_{j_p}^{(+)}$  with $j_{p}=1p_{1/2},1d_{5/2},2s_{1/2},1d_{3/2},1f_{7/2}$,
$2p_{3/2}$, and 2) the s.p.e. of these  $7$ states 
 and the pairing strength, that appear in the 
BCS  gap equation \rf{3.4}, \rf{3.12}-\rf{3.14}, were varied in a $\chi^2$
search in order to account for the experimental spectra $E^{\rm
exp}_j$
 \cite{Ajz85}.
In the PBCS case we proceed in the same way, \ie we solve the
Eqs. \rf{3.4}-\rf{3.6},  and, instead of fitting  
$E_{j_p}^{(+)}$ and $E_{j_h}^{(-)}$ to  $E^{\rm exp}_j$ we fit now  
$ \varepsilon^{Z,N}_{j_p}$ and 
$- \varepsilon^{Z-2,N-2}_{j_h}$.
We have  considered the faraway orbitals $1s_{1/2}$, $2p_{1/2}$ and
$1f_{5/2}$ as well, assuming the first one to be a pure hole state and
the other two pure particle states. Their s.p.e. were taken to by that of
a harmonic oscillator (HO) with standard parameterization \cite{Sie87}.
The single-particle wave
functions were also approximated with that of the HO with the length parameter $b=1.67$
fm, which corresponds to the estimate $ \hbar \omega
=45A^{-1/3}-25A^{-2/3}$~ MeV for the oscillator energy.
The final results for neutrons are shown 
in Table \ref{TableIII} and  those for protons  in  Table \ref{TableIV}.
It is worth noting that  the PBCS   neutron-energy
$\varepsilon^{N}_{1p_{3/2}}=-1.28$ MeV  nicely agrees
with the experimental energy $E_{3/2^-_1}=-1.26$ MeV 
in $^{13}C$. In the same way the PBCS 
proton-energy $\varepsilon^{Z}_{1p_{3/2}}=1.46$ MeV agrees
with the measured energy $E_{3/2^-_1}=1.55$ MeV in $^{13}N$ \cite{Ajz85}.
This does not happen in the BCS case, where  $E^{(+)}_{1p_{3/2}}=-5.07$ MeV 
for neutrons and  $E^{(+)}_{1p_{3/2}}=-2.44$ MeV 
for protons, which clearly shows  the necessity for the number
projection procedure. At this point it could be useful to remember that, while
in $^{11}C$ and  $^{11}B$ the state $3/2^-_1$ is dominantly a hole
state, in  $^{13}C$ and $^{13}N$ it is basically a two-particle-one-hole state.

\begin{figure}[h]
\begin{center}
   \leavevmode
   \epsfxsize = 7cm
     \epsfysize = 8cm
    \epsffile{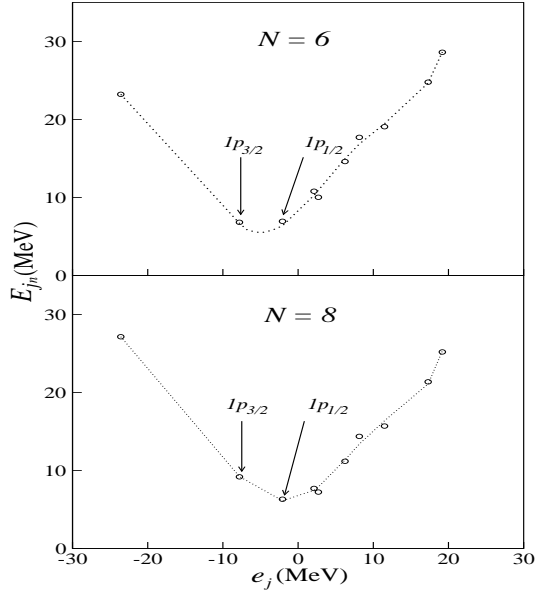}
\end{center}
\caption{\footnotesize
Neutron quasiparticle excitation energies for $N=6$ and $N=8$. 
The states are ordered as  
$1s_{1/2}$, $1p_{3/2}$, $1p_{1/2}$, $2s_{1/2}$, $1d_{3/2}$,
$1f_{7/2}$, $2p_{3/2}$,
$2p_{1/2}$, and $1f_{7/2}$, and the energies are indicated by circles.}
\label{fig1}\end{figure} 


Before proceeding let us remember an important issue
in the description of the $N\cong Z$ nuclei within the QRPA, which is
more inherent to the model itself that to the  occasional parameterization
that might be employed. In fact, a few years ago Volpe \etal
~\cite{Vol00} have  called  attention to the
inconveniences  of applying QRPA to ${^{12}N}$, since
the lowest state turned out not to be the most
collective one. Later on we have shown
\cite{Krm02} that the origin of this difficulty was  the
degeneracy among  the
four lowest proton-neutron
two-quasiparticle states $\ket{1p_{1/2}1p_{3/2}}$,
$\ket{1p_{3/2}1p_{3/2}}$, $\ket{1p_{1/2}1p_{1/2}}$ and
$\ket{1p_{3/2}1p_{1/2}}$. 
 It also has been shown in Ref. \cite{Krm02} that it is imperative to
use the projected QRPA for a physically sound description
of  the weak processes among  the ground states of the triad
$\{{{^{12}B},{^{12}C},{^{12}N}}\}$. 
In fact, when the Fermi level is
fixed at $N=Z=6$, their BCS energies
\br
\E_{j_pj_n}&=& \left\{ \begin{array}{c}
  {E}_{j_p}^{(+)}-{E}_{j_n}^{(-)}=  {E}_{j_p}+{E}_{j_n}
+\lambda_{\rm p} -\lambda_{\rm n};~\mbox{ for}~{^{12}N},  \\
- {E}_{j_p}^{(-)}+{E}_{j_n}^{(+)}=   {E}_{j_p}+{E}_{j_n}
-\lambda_{\rm p} +\lambda_{\rm n};\mbox{ for}~ {^{12}B}, \\
 {E}_{j_p}^{(+)}+{E}_{j_n}^{(+)}= {E}_{j_p}+{E}_{j_n}
+\lambda_{\rm p} +\lambda_{\rm n};~~\mbox{ for}~{^{14}N},  \\
 - {E}_{j_p}^{(-)}-{E}_{j_n}^{(-)}=  {E}_{j_p}+{E}_{j_n}
-\lambda_{\rm p} -\lambda_{\rm n};\mbox{ for}~{^{10}B},
\end{array}  \right.
\label{4.2}\er 
are almost degenerate for all four odd-odd $(Z\pm
1,N\mp1)$ and $(Z\pm 1,N\pm 1)$ nuclei.  As illustrated in
Fig. \ref{fig1}, this, in turn, comes from the fact that 
for $N=Z=6$ the quasiparticle energies $E_{1p_{1/2}}$ and
$E_{1p_{3/2}}$
are very close to  each other.
In the upper panel of Fig. \ref{fig2} are shown the BCS energies \rf{4.2},
 as a function of $N$,
of the just mentioned four states in the case of  nitrogen
isotopes. One notices that for $N\ne Z$  the degeneracy  
is removed yet only partially.
But,  as seen from the lower
panel in Fig. \ref{fig2}, the degeneracy discussed above is totally
removed when the number projection is done.

\begin{figure}[h]
\begin{center}
\vspace{2.cm}
    \leavevmode
     \epsfxsize = 7cm
     \epsfysize = 8cm
     \epsffile{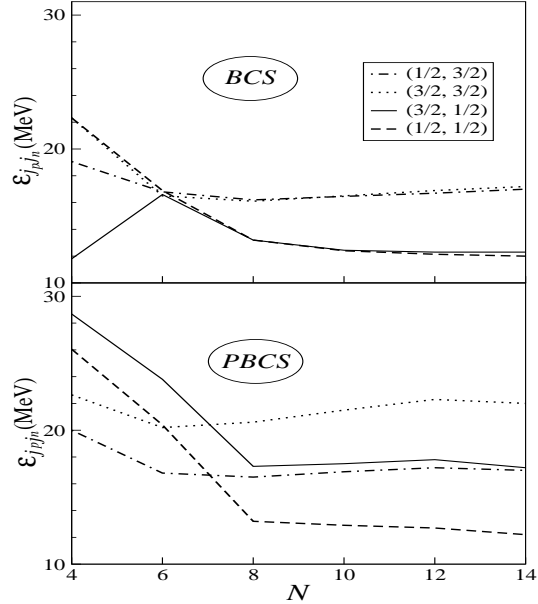}
   \end{center}
\caption{\footnotesize Unperturbed two-quasiparticle energies $\E_{j_pj_n}$ for
 nitrogen isotopes, 
 as a function of $N$,  of the
states $\ket{1p_{1/2}1p_{3/2}}$,
$\ket{1p_{3/2}1p_{3/2}}$, $\ket{1p_{1/2}1p_{1/2}}$ and
$\ket{1p_{3/2}1p_{1/2}}$. In the upper (lower) panel are shown the
 BCS (PBCS) results.} 
\label{fig2}\end{figure}

 
\begin{figure}[h]
\begin{center}
\vspace{2.cm}
    \leavevmode
     \epsfxsize = 7cm
     \epsfysize = 8cm
     \epsffile{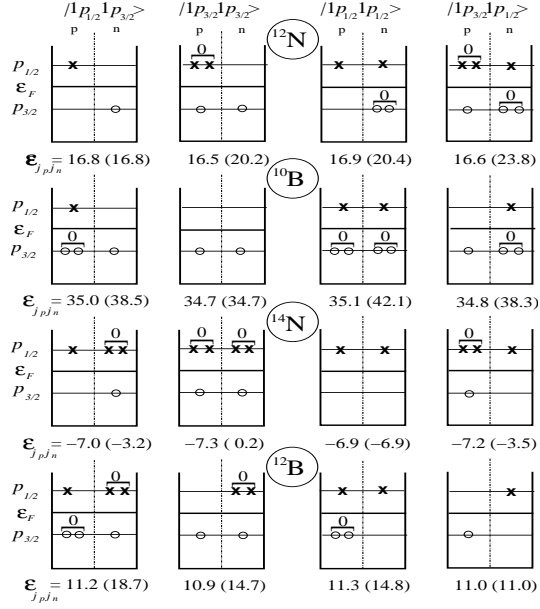}
   \end{center}
\caption{\footnotesize  Schematic representation of the particle-hole 
limits for the seniority-two pn-states. The zero-angular-momentum couplings of 
two-particles or two-holes are indicated by a horizontal bracket.
$\varepsilon_F$ represents the Fermi energy. The  
unperturbed energies $\E_{j_pj_n}$ are given in MeV, being the
PBCS results   given in brackets.}
\label{fig3}\end{figure}

Moreover,  within the PBCS 
 in the case of ${^{12}N}$, for instance, we get 
from numerical calculations 
\br 
\E_{1p_{3/2}1p_{3/2}}&\cong&\E_{1p_{1/2}1p_{1/2}}\cong
\E_{1p_{1/2}1p_{3/2}}+\Delta,
\nn\\
\E_{1p_{3/2}1p_{1/2}}&\cong& \E_{1p_{1/2}1p_{3/2}}+2\Delta,
\label{4.3}\er
with $\Delta=3.4$ MeV.
The meaning of this results can be easily 
disentangled by referring  to the shell-model and  analyzing the 
particle-hole (ph)  limits of the proton-neutron
two-quasiparticle states, which are pictorially shown in Fig. \ref{fig3}. 
One sees that, while $\ket{1p_{1/2}1p_{3/2}}$ corresponds to a
1p1h state in  ${^{12}N}$,
$\ket{1p_{3/2}1p_{3/2}}$ and  $\ket{1p_{1/2}1p_{1/2}}$ correspond to
2p2h states,  and
$\ket{1p_{3/2}1p_{1/2}}$ to a 3p3h state in the same nucleus.
Therefore one can expect that the energy ordering of these states 
would be given by \rf{4.3},  
with  $\Delta=\Delta_{\rm ls}- \Delta_{\rm pair}$ being the energy difference
 between  the spin-orbit splitting, $\Delta_{\rm ls}$, and
the pairing energy, $\Delta_{\rm pair}$.
A similar discussion is pertinent to the remaining three nuclei 
${^{10}B}$, ${^{14}N}$, and ${^{12}B}$. 
That is, one expects that their lowest states would be, respectively,  
$\ket{1p_{3/2}1p_{3/2}}$,  $\ket{1p_{1/2}1p_{1/2}}$ and
$\ket{1p_{3/2}1p_{1/2}}$, as it happens in the PBCS case but not
within the BCS.  Finally, it could be worthwhile to indicate that the 
same pn quasiparticle excitation has quite different ph
compositions in different nuclei. 
This can be observed  by scrutinizing the four columns in
Fig. \ref{fig3}.

\begin{figure}[h]
\begin{center}
   \leavevmode
   \epsfxsize = 7cm
     \epsfysize = 8cm
    \epsffile{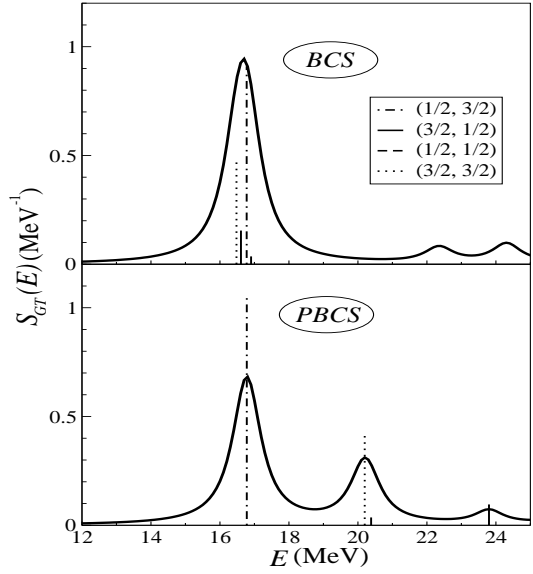}
\end{center}
\caption{\footnotesize The BCS (upper panel) and the PBCS (lower panel) 
Gamow-Teller strength function in $^{12}N$.}
\label{fig4}\end{figure} 

The improvement introduced by the PBCS can also be visualized by making a
direct calculation of the unperturbed GT strength in
$^{12}N$, given by
\br S_{GT}(E)\equiv\frac{1}{\pi}
\sum_{pn}\left| g_A\Bra{p}\s\Ket{n}\right|^2
\frac{\eta}
{\eta^2+(E-\E_{j_pj_n})^2}.
\label{4.4} \er
The BCS and PBCS  results  for  $^{12}N$, when folded with 
$\eta=1$ MeV, are compared in Fig. \ref{fig4}. The PBCS energy
ordering, given by \rf{4.3}, is accompanied by the partial shifting of
the GT strength to higher energies. Therefore it can be said that within the
PBCS the GT resonance is quenched even at the level of the mean field.
 
\begin{figure}[h]
\begin{center}
   \leavevmode
   \epsfxsize = 7cm
     \epsfysize = 8cm
    \epsffile{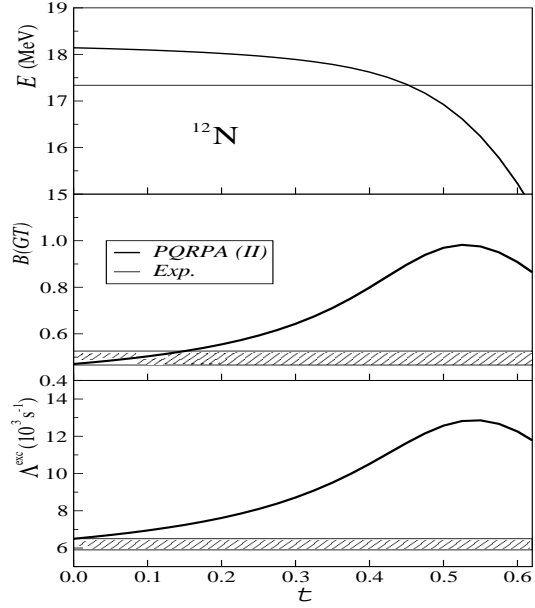}
\end{center}
\caption{\footnotesize The results of the PQRPA calculations, as 
functions of the pp parameter $t$, are confronted with the 
experimental data  taken from Refs. \cite{Ajz85,Al78,Mill72} for:
1) the energy difference $\w_{1^+}$  between the $1^+$ ground state 
in $^{12}N$ and the $0^+$ ground state  in $^{12}C$ (upper panel), 
2) the $B$-value  for  the GT beta transition between these two states
(middle panel), and 
3) the corresponding  muon capture rate  $\Lambda(1^+_1)$ (lower panel). 
The values of $v^{ph}_s$ and $v^{ph}_t$ correspond to    
  PII. }
\label{fig5}\end{figure} 

In view of the above mentioned disadvantages of the standard BCS
approach, from now on we will mainly discuss the number projection
results. In our previous work \cite{Krm02} we have also pointed out that the  
values of the coupling strengths $v_s$ and $v_t$ within the
particle-particle ($pp$) and particle-hole ($ph$) channels
which are  used in $N > Z$ nuclei
($v^{pp}_s\equiv v^{pair}_s$, and $v^{pp}_t\gsim
v^{pp}_s$),  might not be suitable for 
 $N=Z$ nuclei. In fact, the best agreement with data in $^{12}C$
was obtained when the $pp$ channel is totally switched  off,
\ie $v^{pp}_s\equiv
v^{pp}_t=0$, and three different set of values for the ph coupling
strengths \cite{Krm02}, namely:

{\it  Parameterization I} (PI):  $ v^{ph}_s=v^{pair}_s=24$ MeV-fm$^3$, and
 $ v^{ph}_t=v^{ph}_s/0.6=39.86$ MeV-fm$^3$. This means that, the
 singlet ph 
strength is the same as $v^{pair}_s$ obtained from the proton and neutron gap equations,
while the triplet $ph$ depth is estimated from the relation
used by Goswami and Pal \cite{Gos62} in the RPA calculation of ${^{12}C}$.

{\it  Parameterization  II} (PII): $v^{ph}_s=27$  MeV-fm$^3$, and
 $v^{ph}_t=64$ MeV-fm$^3$. These values
were first used  in Refs.  \cite{Nak82,Cas87} and later on in the QRPA
calculations of ${^{48}Ca}$ \cite{Bar98,Hir90a,Krm94}.

{\it  Parameterization  III} (PIII): $v^{ph}_s=v^{ph}_t=45$ MeV-fm$^3$. With
these coupling constants it is possible to reproduce 
fairly well  the energies of the $J^\pi=0_1^+$ and  $1_1^+$ states in
${^{12}B}$ and ${^{12}N}$.


The results displayed in   Fig. \ref{fig5} suggest that
 the choice $v^{pp}_t=0$ for the $pp$ parameter in  the $S=0, T=1$
channel  could be 
appropriate for the description
of the $N=Z$ nuclei. They are shown 
as functions of the parameter
\br
t=\frac{2v_t^{pp}}{v_s^{pair}(p)+v_s^{pair}(n)},
\nonumber \er
together with the experimental data for:
1) the energy difference $\w_{1^+}$  between the $1^+$ ground state 
in $^{12}N$ and the $0^+$ ground state  in
$^{12}C$, 2) the $B$-value  for  the GT beta transition between these
two states, and 3) the corresponding exclusive muon capture rate  
$\Lambda(1^+_1)\equiv\Lambda^{\rm exc}$. 
The values of
$v^{ph}_s$ and $v^{ph}_t$ are that from the PII,
but quite similar results are obtained with other two sets of parameters.
Note that  the $ph$ interaction first 
shifts the energy  $\w_{1^+}$ upwards
by $\sim 1.5$ MeV,  from its unperturbed value
$\E_{1p_{3/2}1p_{3/2}}=16.8$ MeV.  Then, when $t$ is increased,
we have an opposite attractive effect. That is, the $pp$
interaction diminishes $\w_{1^+}$ all up to 
 $t\gtrsim 0.6$, where  the well  known collapse of the QRPA
approximation  occurs.
At variance with what happens in the case of heavy nuclei, here 
 the values of $B(GT)$ and  
$\Lambda(1^+_1)$ basically rise with $t$, and the agreement with the data is
achieved only when the $pp$ interaction is totally switched off.  
Quite generally
the nuclear moments \rf{34} also depend weakly on the
$S=1, T=0$ channel parameter $v^{pp}_s$, for which we adopt as well
the null value,  just to be consistent with our election of  
$v^{pp}_t$.

\begin {table}[h]
\caption {\footnotesize Experimental and calculated muon capture rate 
in units of
$10^3$~s$^{-1}$. The full PQRPA calculations, which include the
relativistic corrections are listed for all three parameterizations
. Theoretical results  that involve only the
velocity-independent matrix elements are displayed 
in parentheses 
in the third column for the case PII. The rates are grouped by
their degrees of  forbiddeness. We show:
i) the  exclusive rates 
 $\Lambda(J^\pi)$, for  $J^\pi_f=1^+_1,1^-_1,2^-_1,2^+_1, $ ii)
the  multipole decomposition of the rates
$\sum_f\Lambda(J^\pi_f)$ for each final state with spin and parity
$J^\pi_f$, and iii)  the  
inclusive decay rate $\Lambda^{\rm
inc}\equiv\sum_{J^\pi_f} \Lambda(J^\pi_f)$.
In column four are listed the results of  recent shell-model
calculations, which are  explained in the text. The measured capture
rates are given in the last column.}
\label{TableV}
\begin{center}
\begin{tabular}{|c|ccc|lc c|c|} \hline
$\mu$ capture& &PQRPA& & &Shell~ Model& &Experiment
\\rate&(I)&(II)&(III)&SM1\cite{Hay00}&SM2\cite{Hay00}&SM3\cite{Aue02}&
\\ \hline 
\hline
\underline{allowed}& &49.3~\%&&~39.3~\%&&&\\
$\Lambda(1^+_1)$&~7.52&6.27(6.50)& 6.27~&~11.56&~6.3&~6.0
&~$6.2\pm0.3$~\cite{Mill72}\\
$\sum_f\Lambda(0^+_f)$&~3.68& 2.86~(3.15)&3.77~&~0.21&&&
\\
$\sum_f\Lambda(1^+_f)$&~20.28&18.14~(18.63)&18.22~&~15.43&&&
\\ 
\underline {1st forbidden}& &46.6~\%&&~55.7~\%&&&\\
$\Lambda(1^-_1)$&~1.06&0.49~(0.51)&0.98~&~& &1.86
&~$0.62\pm0.20$~\cite{Mea01,Sto02}
\\
$\Lambda(2^-_1)$&~0.31&0.18~(0.18)&0.16~&~~&&0.22
&~$0.18\pm0.10$~\cite{Mea01,Sto02}
\\
$\sum_f\Lambda(0^-_f)$&~2.62& 2.35~(0.72)&2.35~&~2.12&&&
\\
$\sum_f\Lambda(1^-_f)$&~11.84&10.37~(9.51)&11.37~&~12.25&&&
\\
$\sum_f\Lambda(2^-_f)$&~7.78& 7.12~(6.90)&7.15~&~7.79&&&
\\
\underline{2nd forbidden}& &3.9~\%&&~4.6~\%&&&\\
$\Lambda(2^+_1)$&~0.19&0.14~(0.16)&0.15~&~~&&0.25
&~$0.21\pm0.10$~\cite{Mea01,Sto02}
\\
$\sum_f\Lambda(2^+_f)$&~1.26& 1.09~(0.89)&1.17~&~1.36&&&
\\
$\sum_f\Lambda(3^+_f)$&~0.63& 0.57~(0.57)&0.58~&~0.46&&&
\\ \hline
$\Lambda^{\rm inc}$&~48.16&42.56~(40.7)&44.67~&~39.82&~41.9&33.5&~$38\pm1$
~~~~\cite{Suz87}
\\\hline
\end{tabular}\end{center}\end {table}

\begin {table}[h]
\caption { \footnotesize Experimental and calculated 
flux-averaged  cross section for the
$^{12}C(\nu_e,e^-)^{12}N$ DAR  reaction in units of $10^{-42}$~cm$^{2}$. 
The full PQRPA calculations, which include the
relativistic corrections, are listed for all three parameterizations. 
Theoretical results  that involve only the
velocity-independent matrix elements are displayed 
in parentheses  
for the case PII in the third column. The multipole decomposition  
$\sum_f\overline{\s}_e(J^\pi_f)$ for each final state with
spin and parity
$J^\pi_f$, as well as the exclusive, 
$\overline{\s}_e^{\rm exc}\equiv\overline{\s}_e(J^{\pi}_f=1^{+}_1) $,
and inclusive  
 $\overline{\s}_e^{\rm inc}=\sum_{J^{\pi}_f}\overline{\s}_e(J^{\pi}_f)$ 
cross sections, are shown. 
In column four are listed the results of a recent shell-model
calculations, which are  explained in the text.  The measured cross
section  are given in the last column.}
\label{TableVI}
\begin{center}
\begin{tabular}{|c|ccc|l c c|c|} \hline
$(\nu_e,e^-)$ cross& &PQRPA&&&Shell ~Model&&Experiment
\\section&(I)&(II)&(III)&SM1\cite{Hay00}&SM2\cite{Hay00}&SM3\cite{Aue02}&
\\ \hline \hline
\underline{allowed}& &83.0~\%&&~83.6~\%&&&\\
$\overline {\s}_e(1^+_1)$
&9.94&8.07~(8.00)&8.17&~20.86 & 7.9 &9.3~ &~$8.9\pm 0.3\pm0.9$~~\cite{Aue01}
\\
$\sum_f\overline{\s}_e(0^+_f) $&1.92 &1.35~(1.31) &2.01 &~0.00&&&
\\
$\sum_f\overline{\s}_e(1^+_f) $&15.98&14.08~(14.22)&12.14&~22.52&&&
\\ 
\underline {1st forbidden}& &16.6~\%&&~16.0~\%&&&\\
$\sum_f\overline{\s}_e(0^-_f) $&0.07 &0.07~(0.05) &0.07 &~0.04&&&
\\
$\sum_f\overline{\s}_e(1^-_f) $&1.94&1.59~(1.43)  &1.80 &~1.90&&&
\\
$\sum_f\overline{\s}_e(2^-_f) $&1.66 &1.43~(1.41)  &1.44 &~2.36&&&
\\ 
\underline{2nd forbidden}& &0.4~\%&&~0.4~\%&&&\\
$\sum_f\overline{\s}_e(2^+_f) $&0.07 &0.05~(0.04) &0.05 &~0.08&&&
\\
$\sum_f\overline{\s}_e(3^+_f) $&0.03 &0.03~(0.03) &0.03 &~0.03&&&
\\ 
\hline
$\overline {\s}_e^{\rm inc}$&
21.67 &18.60~(18.49)&17.54&~26.93&~12.5&15.1
~&~$13.2\pm0.4\pm0.6$~\cite{Aue01}
\\\hline
\end{tabular}\end{center}\end {table}

\begin {table}[h]
\caption { \footnotesize Idem Table \ref{TableVI} but for the 
averaged  exclusive, 
$\overline{\s}_\mu^{\rm exc}\equiv\overline{\s}_\mu(J^{\pi}_f=1^{+}_1) $,
and inclusive  
 $\overline{\s}_\mu^{\rm inc}=\sum_{J^{\pi}_f}\overline{\s}_\mu(J^{\pi}_f)$, 
averaged cross sections for the $^{12}C(\nu_\mu,\mu^-)^{12}N$ DIF reaction
in units of $10^{-40}$ cm$^{2}$.} \label{TableVII}
\begin{center}
\begin{tabular}{|c|ccc|l c c|c|} \hline
$(\nu_\mu,\mu^-)$ cross& &PQRPA& & &Shell ~Model & &Experiment
\\section&(I)&(II)&(III)& SM1\cite{Hay00}&SM2\cite{Hay00}&SM3\cite{Aue02}&
\\ \hline \hline
\underline{allowed}&&20.0~\%&&~17.1~\%&&&\\
$\overline {\s}_\mu(1^+_1)$&
0.74 & 0.59~(0.56)& 0.59&~1.16&~0.56&0.9~&$0.56\pm 0.08\pm0.10$~\cite{Aue02a}
\\
$\sum_f\overline{\s}_\mu(0^+_f) $ &
0.37&0.26~(0.26) &0.39 &~0.11&&&
\\
$\sum_f\overline{\s}_\mu(1^+_f) $ &
2.68&2.33~(2.40) &2.34 &~2.95&&&
\\ 
\underline{1st  forbidden}&&39.5~\%&&~36.4~\%&&&\\
$\sum_f\overline{\s}_\mu(0^-_f) $&
0.03 &0.04~(0.07) &0.04 &~0.07&&&
\\
$\sum_f\overline{\s}_\mu(1^-_f) $&
3.34 &2.79~(2.84) &3.10 &~3.55&&&
\\
$\sum_f\overline{\s}_\mu(2^-_f) $  &
2.53 &2.28~(2.13) &2.29 &~2.91&&&
\\ 
\underline{2nd  forbidden}&&27.1~\%&&~22.2~\%&&&\\
$\sum_f\overline{\s}_\mu(2^+_f) $ &
2.55&2.28~(2.11) &2.37 &~2.59&&&
\\
$\sum_f\overline{\s}_\mu(3^+_f) $&
1.34&1.23~(1.29) &1.23 &~1.39&&&
\\ 
\underline{3th  forbidden}&&10.2~\%&&~17.7~\%&&&\\
$\sum_f\overline{\s}_\mu(3^-_f) $&
0.74 &0.69~(0.73) &0.71 &~1.77&&&
\\
$\sum_f\overline{\s}_\mu(4^-_f) $&
0.68 &0.63~(0.63) &0.63 &~1.41&&&
\\ 
\underline{4th  forbidden}&&3.1~\%&&~6.5~\%&&&\\
$\sum_f\overline{\s}_\mu(4^+_f) $ &
0.22 &0.21~(0.20) &0.21 &~0.66&&&
\\
 $\sum_f\overline{\s}_\mu(5^+_f) $&
0.20 &0.19~(0.19) &0.19 &~0.51&&&
\\\hline
$\overline {\s}_\mu^{\rm inc}$&
14.69&12.94~(12.86)&13.51&~17.92&~13.8&19.2~&$10.6\pm0.30\pm 1.80$~\cite{Aue02a}
\\\hline
\end{tabular}\end{center}\end {table}


Results for the  muon capture rates, and  the neutrino 
$(\nu_e,e^-)$ DAR, and $(\nu_\mu,\mu^-)$ DIF reaction  flux-averaged cross sections are
shown, respectively, in  Tables \ref{TableV}, Tables \ref{TableVI} and
Tables \ref{TableVII}. 


The flux-averaged cross section is defined as
\be
\overline{\s}_\ell(J_{f})= \int_{\Delta_{J_f}} dE_{\nu}
\s_\ell(E_{\nu},J_{f}) {\Phi}_\ell(E_{\nu});~ ~ \ell=e,\mu,
 \label{4.5}\ee
where $\Phi_\ell(E_{\nu})$ is the normalized  neutrino flux. For
electron neutrinos this flux was approximated by the Michel spectrum, and
for the muon neutrinos we used that from Ref. \cite{LSND}. The energy
integration is carried out in the DAR  interval
$m_e+\w_{J_{f}}\le \Delta_{J_{f}}^{\rm DAR}\le 52.8 $ MeV for electrons
and in the  DIF   interval
$m_\mu+\w_{J_{f}}\le \Delta_{J_{f}}^{\rm DIF}\le 300 $ MeV for muons. 

The full PQRPA calculations, which include the
relativistic corrections are listed for all three parameterizations 
I, II, and III. On the contrary, the theoretical results  involving
just the
velocity-independent operators ${\sf Y}_{J}(\k\rb)$  and 
$ {\sf S}_{JL}(\k\rb)$ are displayed 
only for the case PII.
Contributions of the other two operators, ${\sf P}_{JL}(\k\rb)$
and $ {\sf Y}_{J}(\k\rb,\mbs\cdot\vb)$, to the muon capture rates 
are small (of the order of $5\%$)  as displayed in 
Table \ref{TableV}. The only exception are the $0^-$
states where the relativistic operator  $ {\sf
Y}_{0}(\k\rb,\mbs\cdot\vb)$ dominates over the non-relativistic 
one $ {\sf S}_{01}(\k\rb)$. We also see from Tables \ref{TableVI} and
\ref{TableVII} that the nonlocality effects on the neutrino-nucleus
reactions are of the order of $1\%$ and therefore they can be
neglected.

The numerical results are sorted  
according to the order of forbiddeness 
of the transition moments, which can be:
 {\em allowed } (A): $J^\pi=0^+,1^+$, {\em first forbidden} (F1):
$J^\pi=0^-,1^-,2^-$,
{\em second forbidden} (F2): $J^\pi=2^+,3^+$, {\em third forbidden}
(F3): $J^\pi=3^-,4^-$, and so  on.
The response of
the three weak processes to successive multipoles 
is strongly correlated with the average momentum
transfer, $\overline{\kappa}$,  involved in each:
$(\nu_e,e^-)$-reaction, 
$\overline{\kappa}\sim 0.2$ fm$^{-1}$;
 $\mu$-capture, $\overline{\kappa}\sim 0.5$ fm$^{-1}$; and 
 $(\nu_\mu,\mu^-)$-reaction, $\overline{\kappa}\sim 1$ fm$^{-1}$.
As a consequence,  in
the first case the A-moments are by far the dominant ones,
contributing with $\sim
83.0\%$, with  the remaining 
part of the reaction strength  carried almost entirely $(\sim 16.6
\%)$ by the F1-moments.
 For $\mu$-capture, the A- and F1-matrix elements contribute,
respectively, with $49.3\%$ and $46.6\%$, while  
the F2-moments carry  only $3.9\%$ of the total transition rate, and 
the contribution of the F3-ones are negligibly small. Finally,
the inclusive cross
section in  the $(\nu_\mu,\mu^-)$-reaction is spread out rather 
 uniformly over several multipoles and   it is necessary to include 
up to fourth forbidden moments, with their intensities distributed
as follows: $20.0\%$ (A), $39.5\%$ (F1), $27.1\%$ (F2),  $10.2\%$ (F3),  and
 $3.1\%$ (F4).
The exclusive contributions, coming 
from the ground state $1^+_1$, are quite different in the three
cases  due to the just mentioned implication of the momentum 
transfer. That is, of the total transition rates for 
the $(\nu_e,e^-)$-reaction, $\mu$-capture, and
$(\nu_\mu,\mu^-)$-reaction, the exclusive contributions
are, respectively, $43\%$,  $15\%$, and  $5\%$.


Let us say a few words on the comparison of our results with the
experimental data: 
\bit 
\item {\em $\mu$ capture} (Table \ref{TableV}): All    exclusive rates 
 $\Lambda(J^\pi)$ with   $J^\pi_f=1^+_1,1^-_1,2^-_1,2^+_1$, are fairly
 well accounted for by the theory, while the inclusive rate,  
$\Lambda^{\rm inc}$, is overpredicted by $\sim 10\%$. 
\item $(\nu_e,e^-)$-reaction (Table \ref{TableVI}): Even though
the exclusive cross section, $\overline{\s}_e^{\rm exc}$, is well reproduced
within the PQRPA, the inclusive one,  $\overline{\s}_e^{\rm inc}$,
 is $\sim 40\%$ above the data.
A plausible explanation for this difference could be the fact that
we find a  very significant  amount of the GT strength ($32\%)$ and the
Fermi strength ($7\%)$  within the DAR energy interval,
$ 18$  MeV $\lsim E_\nu \lsim 50$ MeV, where 
 the electron-neutrino 
flux, $\Phi_e(E_{\nu_e})$, changes  very abruptly,
making  the inclusive cross section  to be very sensitive to the strength
distribution of low-lying excited states.

\item $(\nu_\mu,\mu^-)$-reaction (Table \ref{TableVII}):
Here also the exclusive cross section, $\overline{\s}_\mu^{\rm exc}$,
is in full  agreement with the data, whereas the inclusive one, 
$\overline{\s}_\mu^{\rm inc}$, is overpredicted by  $\sim 20\%$. 

 \eit

In the last three tables we also confront our  PQRPA results with the
shell model calculations performed by: a) Hayes and Towner \cite{Hay00},
within the model spaces called by them as (iii) and (iv), and 
which are labeled here,  respectively, as
SM1 and SM2, and b) Auerbach and Brown \cite{Aue02} which we label as SM3.
The multipole breakdown in the contributions to the cross sections
from each multipole is only given  for the SM1 scheme 
\cite{Hay00}.   
At first glance our results seem 
to agree fairly well with the shell model ones, particularly when 
the amounts of  allowed  and forbidden transition strengths  
are confronted.  However, this is not true,  
as can be seen from a more careful 
analysis of the multipole structure of the transition rates.

For instance, 
 the total positive 
parity capture  rates with the
$ 1^+_1$ state  excluded, \ie  $\sum_{J_f\ne 1^+_1  }\Lambda(J^+_f)$, 
are equal to $5.9$ and $3.6$ in SM1 and SM3,
 respectively, while we get $16.4$ (in units of $10^3$ s$^{-1}$). 
Note also that in shell model calculations  almost all GT strength
is exhausted by  the ground state transition, 
while within the PQRPA only $35 \%$ of this strength goes into the $1^+_1$
 state.
Another important  discrepancy is in the Fermi transitions, for which we get 
$\sum_f\Lambda(0^+_f)$= 2.86, while  they obtain only  
$\sum_f\Lambda(0^+_f)=0.21$ (in units of
$10^3~s^{-1}$).

One  sees that our $\overline{\s}_e^{\rm exc}$ is consistent with
the SM2 and SM3 shell model calculations, while our $\overline{\s}_e^{\rm inc}$
 falls in between the calculations SM1
and SM2 and is $\sim 20\%$ larger than the SM3 result.
However, same as in the case of
muon capture, we find  much more  $\overline{\s}_e(1^+)$-strength in the excited states of 
$^{12}N$ nucleus than appears in shell model calculations. Namely, we
obtain that
  $\sum_{J_f\ne 1^+_1}  \overline{\s}_e(J^+_f)=7.44$,  
while in  SM1 and SM3 this quantity is, respectively,
 $1.77$ and $0.40$ (in units of $10^{-42}$ cm$^2$). 
More, in all three shell model calculations more than $90\%$
of the $\overline{\s}_e(1^+)$-strength is concentrated in the 
$^{12}N$ ground state meanwhile we find only $57\%$.

In the same way as in the  shell model calculations, within 
the PQRPA  the total DIF cross section  is mainly built 
up from  forbidden excitations. 
Nevertheless, although the  PQRPA results 
agree  with the SM2 calculation, they are quite small when
compared with those provided by the SM1 and SM3 models, for  
both $\overline{\s}_\mu^{\rm exc}$ and 
$\overline{\s}_\mu^{\rm inc}$. The differences 
in $\overline{\s}_\mu^{\rm inc}$ come not 
only from  the positive parity contributions but also from 
the negative parity ones.

As a corollary of the above discussion we would like to stress  that 
 it is not easy to asses  whether the PQRPA results are better or worse
than the shell model ones. We just can say that with the use of
only a few essentially phenomenological parameters  the PQRPA 
is able to account for a large number of weak processes in
$^{12}C$. (The experimental energies of the  ${3/2^-_1}$ state in
$^{13}C$ and $^{13}N$ are also predicted within the PBCS.)
  That is, the s.p.e., $e_j$,  for $j=1p_{3/2},1p_{1/2},
1d_{5/2},2s_{1/2},1d_{3/2},1f_{7/2},2p_{3/2}$, and the pairing strength
$v_{\sss{s}}^{\sss{pair}}$ have been fixed 
from  of the experimental energies of the  odd-mass nuclei  
$^{11}C$, $^{11}B$, $^{13}C$ and $^{13}N$, while the particle-hole
coupling  strengths  $v^{ph}_s$  and
 $v^{ph}_t$ were taken from the previous QRPA calculations of 
${^{48}Ca}$ \cite{Bar98,Hir90a,Krm94}.
Only the particle-particle couplings $v^{pp}_s$
and $v^{pp}_t$ has been treated as free parameters,  
and we have used here   $v^{pp}_s=v^{pp}_t=0$.
It  could be interesting to inquire  whether, this 
rather extreme  parameterization  is also  appropriate 
for the description of  other light  $N=Z$ nuclei, such as  $^{14}N$
and $^{16}O$,  which  have been discussed recently within a shell model
scheme \cite{Aue02}.

Finally, we would like to point out that in the  DIF neutrino 
oscillation search \cite{Ath98} an excess signal of 
 \[
N^{\rm exp}_{\nu_\mu\go\nu_e}=18.1\pm6.6\pm 4.0 
\]
events has been observed in the  $^{12}C(\nu_e,e^-)^{12}N$ reaction, 
which are  evaluated theoretically by the  expression 
 \[
N^{\rm th}_{\nu_\mu\go\nu_e}= \int_{\Delta^{\rm osc}} dE_{\nu}
\s_e(E_{\nu})P_{\nu_\mu\go \nu_e}(E_{\nu}) {\Phi}_\mu(E_{\nu});~
~~~\s_e(E_{\nu})\equiv  \sum_{J_f}\s_e(E_{\nu},J_{f}), 
\]
where $77.3$ MeV $\le \Delta^{\rm osc}\le 217.3$ MeV 
stands for   the experimental energy window. The oscillation 
probability reads
\[
P_{\nu_\mu\go \nu_e}(E_{\nu})=
\sin^2(2\theta)\sin^2\left(\frac{1.27 \Delta m^2 L}{E_\nu}\right),
\] 
where $\theta$ is the mixing angle between the neutrino mass  
eigenstates, $\Delta m^2$ is the difference in neutrino
eigenstate masses squared, in eV$^2$, and $L$ is the distance in meters
traveled by the neutrino from the source. One sees therefore that the
extraction
of the permitted values of $\theta$ and $\Delta m^2$ from experimental
data might depend critically on 
the theoretical estimate of $\s_e(E_{\nu})$. 
So far, the 
electron cross section obtained  within  the
continuum random phase approximation (CRPA) \cite{Kol99}
has been used  \cite{Ath98}. 
This $\s_e(E_{\nu})$ is confronted in  Figure \ref{fig6} 
with our QRPA and PQRPA results calculated  with PII. 
As can be noticed the PQRPA yields a substantially different  
$\s_e(E_{\nu})$, 
inside the experimental energy  window for the neutrino energy. 
The consequences  of this difference on the confidence regions for 
$\sin^2(2\theta)$ and $\Delta m^2$ will be discussed in a future 
 paper.


\begin{figure}[t]
\begin{center}
   \leavevmode
   \epsfxsize = 9cm
     \epsfysize = 8cm
    \epsffile{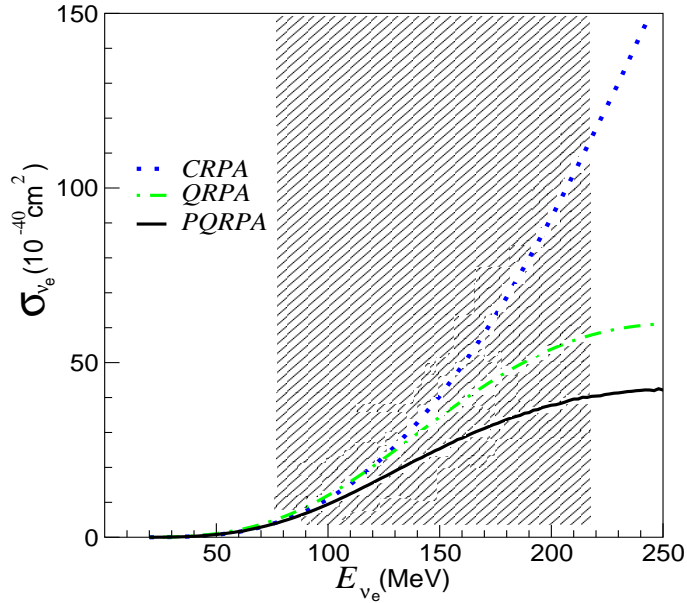}
\end{center}
\vspace{1cm}
\caption{\footnotesize (Color online) Calculated cross section
 $\s_{\nu_e}$ as function of the neutrino energy. The dashed 
region indicates the experimental energy  window.}
\label{fig6}\end{figure} 

After finishing this work we have learned that quite recently
Nieves \etal~ \cite{Nie04} were able to describe rather well 
the inclusive muon capture rate in $^{12}C$,  and the inclusive  
$^{12}C(\nu_\mu,\mu^-) ^{12}N$ and $^{12}C(\nu_e,e^-) ^{12}N$ 
cross sections, 
within the framework of a Local
Fermi Gas picture which includes the RPA correlations.

\begin{center}
{\bf ACKNOWLEDGEMENTS}
\end{center}
The authors wish to express his sincere thanks to Tatiana Tarutina
for the meticulous reading of the manuscript. One of us (A.S.) 
acknowledges support received from the CLAF-CNPq Brazil.

\newpage



\begin{thebibliography}{9}

\bibitem{Ath98} LSND Collaboration, C. Athanassopoulos
\etal, 
Phys. Lett. {\bf 81}, 1774 (1998); Phys. Rev. C {\bf 58}, 2489 (1998).

\bibitem {Agu01} LSND Collaboration, A. Aguilar \etal, 
Phys. Rev. D {\bf 64}, 112007 (2001).

\bibitem{Eit99} KARMEN Collaboration, K. Eitel \etal, 
Nucl. Phys. B, Proc.Suppl. {\bf 77}, 212 (1999).

\bibitem{Aue01} LSND Collaboration, L-B. Auerbach \etal , Phys. 
Rev. C {\bf 64}, 065501 (2001).

\bibitem{Aue02a} LSND Collaboration, 
L-B. Auerbach \etal, 
 Phys. Rev. C {\bf 66}, 015501 (2002).

\bibitem{Don79} T.W. Donnelly and W.C. Haxton,
Atomic Data and Nuclear Data Tables {\bf 23}, 103 (1979); T.W. 
Donnelly and R. D. Peccei, Phys. Rep. {\bf 50}, 1 (1979).

\bibitem{Wal95} J.D. Walecka, {\it Theoretical Nuclear and Subnuclear Physics},
{\it Oxford University Press, New York}, 531 (1995).

\bibitem{Aue97} N. Auerbach, N. Van Giai, and O.K. Vorov,  
Phys. Rev. C {\bf 56}, R2368 (1997).

\bibitem{Vol00} C. Volpe, N. Auerbach, G. Col\`o, T. Suzuki, and N. Van
Giai, Phys. Rev. C {\bf 62}, 015501 (2000).

\bibitem{Kur90} T. Kuramoto, M. Fukigita, Y. Kohyama and K. Kubodera,
Nucl. Phys. A {\bf 512}, 711 (1990).

\bibitem{Luy63} J.R.Luyten, H.P.C. Rood and H.A. Tolhoek,
  Nucl.Phys. {\bf 41}, 236 (1963).

\bibitem{Aue83} N. Auerbach and A. Klein,  Nucl. Phys. A {\bf 395}, 77 (1983).

\bibitem{Bar98} C. Barbero, F. Krmpoti\'c, and D. Tadi\'c,
Nucl. Phys. A {\bf 628}, 170 (1998);
 C. Barbero, F. Krmpoti\'c, A. Mariano and D. Tadi\'c, Nucl. Phys.
A {\bf 650}, 485 (1999).

\bibitem{Krm02} F. Krmpoti\'c, A. Mariano, and A. Samana,  
Phys. Lett. B {\bf 541}, 298 (2002).

\bibitem{Sup64} I. Supek, {\it Teorijska Fizika i Struktura Materije},
(Zagreb), Vol. II  (1964).


\bibitem{Hay00} A.C. Hayes and I.S. Towner,
Phys. Rev. C {\bf 61}, 044603 (2000).

\bibitem{Aue02} N. Auerbach and  B.A. Brown, 
Phys.  Rev. C {\bf 65}, 024322 (2002).

\bibitem{Bli66} R.J. Blin-Stoyle and S.C.K. Nair, 
Adv. Phys. {\bf 15},  493 (1966).

\bibitem{Bro85} B.A. Brown and B.H. Wildenthal, At. Data Nucl. Data Tables
{\bf 33}, 347 (1985).

\bibitem{Cas87} H. Castillo and F. Krmpoti\'c, Nucl. Phys. {\bf A 469}, 637
(1987).
\bibitem{Ost92} F. Osterfeld, Rev. Mod. Phys. {\bf 64}, 491 (1992).


\bibitem{Tow95}I.S. Towner and  J.C. Hardy, 
{\it The Nucleus as a Laboratory for
Studying Symmetries and Fundamental Interactions}, eds. E.M.
Henley and W.C. Haxton, nucl-th/9504015.

\bibitem{Krm93} F. Krmpoti\'c, A. Mariano, T.T.S. Kuo, and
 K. Nakayama,   Phys. Lett. B {\bf 319}, 393 (1993).

\bibitem{Hir90a} J. Hirsch and F. Krmpoti\'c,
Phys. Rev. C {\bf 41}, 792 (1990).

\bibitem{Hir90b} J. Hirsch and F. Krmpoti\'c,
Phys. Lett. B {\bf 246}, 5 (1990).

\bibitem{Krm92} F. Krmpoti\'c, J. Hirsch and H. Dias,
Nucl. Phys. A {\bf 542}, 85 (1992).

\bibitem{Ajz85} F. Ajzenberg-Selove, Nucl. Phys. A {\bf 433}, 1 (1985)
; TUNL Nuclear Data Evaluation Project, available 
WWW: http:// www.tunl.duke.edu/nucldata/

\bibitem{Sie87} P.J. Siemens and A.S. Jensen, {\it Elements of
Nuclei: Many Body Physics with the Strong Interction} 
(Addison-Wesley Publishing Company Inc., Redwood City, California, 1987).


\bibitem{Gos62} A. Goswami and M.K. Pal, Nucl. Phys. {\bf 35}, (1962) 544.

\bibitem{Nak82} K. Nakayama, A. Pio Gale\~{a}o and F. Krmpoti\'{c},
Phys. Lett. {\bf B 114} (1982) 217.

\bibitem{Krm94} F. Krmpoti\'c and Shelly Sharma,
 Nucl. Phys. {\bf A 572}, (1994) 329.

\bibitem{Al78} D. E. Alburger and A.M. Nathan, Phys. Rev. {\bf C 17}, 
(1978) 280.

\bibitem{Mill72} G. H. Miller \etal , Phys. Lett. B {\bf 41}, 50 (1972).


\bibitem{Mea01} D.F. Measday, Phys. Rep. {\bf 354}, 243 (2001).

\bibitem{Sto02}T.J. Stocki, D.F. Maesday, E. Gete, M.A. Saliba, and
T.P. Gorrinde, Nucl. Phys. A {\bf 697}, 55 (2002).

\bibitem{Suz87} T. Suzuki \etal ,
Phys. Rev. C {\bf 35}, 2212 (1987).


\bibitem{LSND} LSND home page,
http://www.nu.to.infn.it/exp/all/lsnd/


\bibitem{Kol99} E. Kolbe, K. Langanke, and P. Vogel,
Nuc. Phys. A {\bf 652}, 91 (1999).

\bibitem{Nie04}
J. Nieves, J.E. Amaro, and M. Valverde,  Phys. Rev. C, in press,
nucl-th/0408005, {\it ibidem}  nucl-th/0408008;
J.E. Amaro, C. Maieron, J. Nieves, and M. Valverde,
nucl-th/0409017.













\end{thebibliography}
\end{document}